\documentclass[tighten,twocolappendix,twocolumn]{aastex631}

\usepackage[T1]{fontenc}
\usepackage{amsmath}
\usepackage{xspace}
\usepackage{enumitem}

\graphicspath{{figures/}}

\newcommand{\ud}{\mathrm{d}}
\newcommand{\hodgestar}{\raisebox{1pt}{$\star{}$}}

\newcommand{\bhspin}{a_*}
\newcommand{\sgra}{{Sgr A$^*$}\xspace}
\newcommand{\ipole}{{\tt ipole}\xspace}
\newcommand{\kharma}{{KHARMA}\xspace}
\newcommand{\koral}{{\tt KORAL}\xspace}

\defcitealias{chael_2023_bhp1}{Paper I}

\begin{document}

\shorttitle{}
\shortauthors{}

\title{Black Hole Polarimetry II: The Connection Between Spin and Polarization}

\author[0000-0001-6952-2147]{George N.~Wong}
\correspondingauthor{George N.~Wong}
\email{gnwong@ias.edu}
\affiliation{Princeton Gravity Initiative, Princeton University, Princeton, New Jersey 08544, USA}
\affiliation{School of Natural Sciences, Institute for Advanced Study, 1 Einstein Drive, Princeton, NJ 08540, USA}

\author[0000-0003-2966-6220]{Andrew Chael}
\affiliation{Princeton Gravity Initiative, Princeton University, Princeton, New Jersey 08544, USA}

\author[0000-0002-1559-6965]{Alexandru Lupsasca}
\affiliation{Department of Physics \& Astronomy, Vanderbilt University, Nashville, TN 37212, USA}

\author[0000-0001-9185-5044]{Eliot Quataert}
\affiliation{Princeton Gravity Initiative, Princeton University, Princeton, New Jersey 08544, USA}
\affiliation{Department of Astrophysical Sciences, Princeton University, Princeton, NJ 08544, USA}

\begin{abstract}
We study synchrotron polarization in spatially resolved horizon-scale images, such as those produced by the Event Horizon Telescope (EHT). In both general relativistic magnetohydrodynamic (GRMHD) simulations as well as simplified models of the black hole magnetosphere, the polarization angle, quantified by the complex observable $\angle\beta_2$, depends strongly and systematically on the black hole spin. This relationship arises from the coupling between spin and the structure of the magnetic field in the emission region, and it can be computed analytically in the force-free limit. To explore this connection further, we develop a semi-analytic inflow framework that solves the time stationary axisymmetric equations of GRMHD in the black hole's equatorial plane; this model can interpolate between the force-free and inertial regimes by varying the magnetization of the inflow. Our model demonstrates how finite inertia modifies the structure of the electromagnetic field and can be used to quantitatively predict the observed polarization pattern. By comparing reduced models, GRMHD simulations, and analytic limits, we show that the observed synchrotron polarization can serve as a robust diagnostic of spin under assumptions about Faraday rotation and the emission geometry. Applied to EHT data, the model disfavors high-spin configurations for both M87$^*$ and \sgra, highlighting the potential of polarimetric imaging as a probe of both black hole spin and near-horizon plasma physics.
\end{abstract}


\section{Introduction}

The Event Horizon Telescope (EHT) has produced the first spatially resolved polarimetric images of emission near the horizons of supermassive black holes. These images reveal bright, compact emission rings threaded by coherent, helical polarization patterns. In both Sagittarius A* (\sgra), the Milky Way’s central black hole, and M87$^*$, the black hole in the nearby giant elliptical galaxy M87, the EHT polarimetric images exhibit high linear polarization fractions and organized azimuthal polarization on scales of only a few gravitational radii 
\citep{eht_m87_7,eht_m87_8,eht_sgra_7,eht_sgra_8}. These signatures have been interpreted as evidence for strong, ordered magnetic fields near the event horizon. Such fields play a central role in coupling black holes to their environments through the Blandford-Znajek mechanism (BZ; \citealt{blandford_1977_bz}), in which horizon-scale magnetic flux is wound up by the black hole’s rotation, enabling the extraction of spin energy. The resulting Poynting flux can launch and collimate relativistic outflows, establishing a natural pathway for transporting rotational energy from the black hole to large-scale jets \citep[e.g.,][]{eht_m87_8}. In this way, the EHT polarization results not only reveal the character of the horizon-scale magnetic field, but also connect directly to the long-standing question of how relativistic jets are powered and sustained \citep{chael_2023_bhp1}.

One of the most useful tools for analyzing polarimetric black hole images is the complex coefficient $\beta_2$, which quantifies the net rotational symmetry of the electric vector position angle across an image \citep{palumbo_2020_beta2}. The image-integrated value of $\angle \beta_2$ has been shown to provide a robust observable that encodes information about the magnetic field geometry and energy flow near the black hole. Simulated images produced from general relativistic magnetohydrodynamic (GRMHD) simulations support this connection. Synchrotron radiation emitted from turbulent plasma orbiting a spinning black hole naturally produces polarized structures whose distributions of $\angle \beta_2$ are correlated with the dimensionless spin parameter $\bhspin = J c / G M^2$ \citep{palumbo_2020_beta2,himwich_2020_universalpol,gelles_2021_polmidplane,eht_m87_8,ricarte_2023_resolvedpol,eht_sgra_8}. This trend arises from a combination of radiative transfer effects, general relativistic propagation, and the increasing azimuthal twist of magnetic field lines at higher spin. Nevertheless, while the correlation between $\angle \beta_2$ and $\bhspin$ is clear, its physical origin and robustness remain incompletely understood, leaving open the challenge of establishing $\angle \beta_2$ as a definitive tracer of black hole spin.

This paper is the second in a series investigating how polarimetric observables in black hole images connect to the underlying structure of the magnetosphere. In Paper~I~\citep{chael_2023_bhp1} we showed that $\angle \beta_2$ provides a direct probe of the direction of electromagnetic energy flux near the black hole. In this work, we investigate how plasma dynamics outside the event horizon shapes the polarization structure observable in black hole images. While previous studies have shown that near-horizon magnetic fields encode information about spin and energy extraction, the effects of plasma inertia on the polarization pattern remain poorly understood. We address this question by analyzing a spectrum of models for the black hole magnetosphere ranging from the idealized BZ force-free monopole to fully dynamical GRMHD simulations. To interpolate between the two extremes, we introduce a stationary inflow model that differs from previous semi-analytic models (see \citealt{broderick_2006_riaf,pu_2018_riafsgra,ozel_2022_bhgrtests,cardenasavendano_2023_aart}) by directly solving the time-stationary axisymmetric GRMHD equations in the equatorial plane. The model incorporates plasma inertia through a tunable magnetization parameter. The model is analytically tractable, computationally inexpensive, and we show that it captures the spin-dependent polarization trends observed in numerical simulations.

The inflow model offers a simplified framework for studying how fluid motion and the geometry of the magnetic field together shape the observable polarization pattern. We use the model to investigate how plasma properties like the degree of field winding, the inflow velocity, and the effective magnetization affect both the image-integrated value of $\angle \beta_2$ and its radial profile. The model can interpolate between the two well-studied limits of the idealized force-free BZ solution and the time-averaged GRMHD simulations of strongly magnetized accretion flows. This interpolation allows us to identify the physical mechanisms that cause numerical simulations to deviate from their force-free counterparts and to isolate how plasma inertia modifies the polarization signatures of spin-driven magnetospheres.

We show that the previously identified spin dependence of the polarization pattern is governed by a small set of geometric and dynamical quantities tied to the large-scale electromagnetic field, rather than by detailed microphysical modeling of turbulence or radiation. In this way, the inflow model functions as a fast and interpretable surrogate for GRMHD, enabling rapid exploration of parameter space while clarifying the theoretical origin of the observed trends. We further demonstrate how the model can be used to inform constraints on black hole spin in sources like M87$^*$ and \sgra.

The remainder of this paper is organized as follows. In Section~\ref{sec:bhpolarimetry}, we review the definition of $\beta_2$ and its physical interpretation. Section~\ref{sec:models} introduces models for the black hole magnetosphere and describes the inflow model in detail. In Section~\ref{sec:model_comparison}, we compare the predictions of the inflow model with those from an example GRMHD simulation, and in Section~\ref{sec:polarization_spin} we show inflow model predictions for $\angle \beta_2$ across spin. We conclude in Section~\ref{sec:discussion} with a discussion of implications for EHT observations, limitations of the model, and prospects for future work.

\section{Polarimetric Observables}
\label{sec:bhpolarimetry}

Polarized light from synchrotron radiation can be used as a direct probe of the magnetic field geometry near the black hole event horizon, and the ordered patterns observed in EHT images suggest underlying rotational symmetries in the field. In this section, we introduce the $\beta_m$ formalism, focusing on the $m=2$ mode that describes the dominant structure in M87$^*$ and \sgra.

\begin{figure*}[htp!]
\centering
\includegraphics[width=\textwidth]{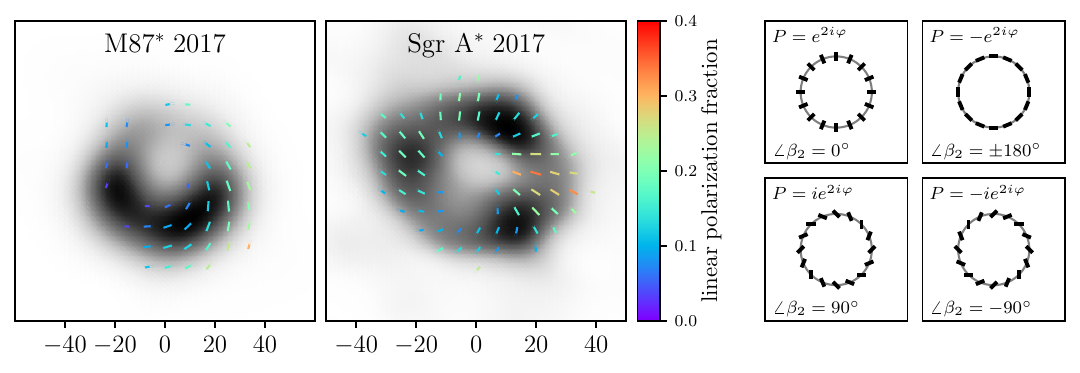}
\caption{
Polarimetric structure of M87$^*$ and \sgra as observed by the EHT. Left and center: Polarimetric images of M87$^*$ and \sgra from the 2017 Event Horizon Telescope campaign (\citealt{eht_m87_7,eht_sgra_7}; M87$^*$ shown on April 11, day 3601). Greyscale indicates total intensity; tick marks show the direction and relative magnitude of linear polarization, with color denoting fractional polarization. Both sources exhibit a coherent spiral near the black hole, signaling a strong $m = 2$ mode in the polarization field.
Right: Schematic of the complex polarization mode $\beta_2$, which quantifies the amplitude and handedness of the azimuthal $m = 2$ component. The magnitude $|\beta_2|$ reflects the strength of the mode, while the phase $\angle\beta_2$ encodes its orientation. Purely radial and azimuthal fields correspond to $\beta_2 = \pm 1$, and left- or right-handed spirals to $\beta_2 = \pm i$. For M87$^*$, $\angle\beta_2$ ranges from $-163^\circ$ to $-129^\circ$, and for \sgra, Faraday-corrected values span between $-168^\circ$ and $-85^\circ$.
}
\label{fig:ehtobs_beta2}
\end{figure*}

\subsection{Quantifying rotational symmetry in black hole images}

Polarimetric black hole images frequently display coherent, spiral-like patterns in the electric vector position angle (EVPA). These features are especially prominent for nearly face-on viewing geometries, as inferred for M87$^*$ from the position angle of the large-scale jet and from VLBI observations \citep{walker_2018_m87jet,eht_m87_5} as well as more tentatively for \sgra from VLBI and infrared interferometry \citep[e.g.,][]{eht_sgra_5,eht_sgra_8,gravity_2018_orbit}. Such spiral structures reflect an underlying rotational symmetry in the near-horizon magnetic field and naturally motivate a Fourier-like decomposition of the polarization map to isolate dominant azimuthal modes. This decomposition is conveniently captured by the complex $\beta_m$ coefficients introduced by \citet{palumbo_2020_beta2}, which encode the amplitude and phase of the $m$-th Fourier mode of the polarization field.

The rotationally symmetric linear polarization pattern is quantified with the complex $\beta_2$ coefficient. We work in the polar image-plane coordinates $(\rho, \varphi)$, where $\rho$ is the radial distance from the image center and $\varphi$ is the azimuthal angle measured counterclockwise on the sky. The complex linear polarization is given by $P(\rho, \varphi) = Q(\rho, \varphi) + i U(\rho, \varphi)$, where $Q$ and $U$ are the Stokes parameters. The $m$-th Fourier mode coefficient $\beta_m$ is then computed as
\begin{align}
\beta_m &= \dfrac{1}{I_{\rm ann}} \int\limits_{\rho_{\rm min}}^{\rho_{\rm max}} \int\limits_0^{2\pi} P(\rho, \varphi) e^{-i m \varphi} \; \rho \, d\rho \, d\varphi , \\
I_{\rm ann} &= \int\limits_{\rho_{\rm min}}^{\rho_{\rm max}} \int\limits_0^{2\pi} I(\rho, \varphi) \; \rho \, d\rho \, d\varphi ,
\label{eqn:bzdefn}
\end{align}
where $I(\rho, \varphi)$ is the total intensity and $I_{\rm ann}$ is the total flux within the annulus extending between $\rho_{\rm min}$ and $\rho_{\rm max}$.

The magnitude $|\beta_2|$ characterizes the strength of the rotationally symmetric mode, while the phase $\angle \beta_2$ captures the orientation of that symmetry relative to a purely radial pattern. For example, $\angle \beta_2 = 0$ corresponds to a purely radial polarization field, $\angle \beta_2 = \pi$ to a purely azimuthal field, and intermediate values indicate spirals with left- or right-handed helicity. This is illustrated in Figure~\ref{fig:ehtobs_beta2}, which shows both real EHT polarimetric data and schematic patterns corresponding to different $\beta_2$ values.

Since linear polarization traces the geometry of the magnetic field, the complex coefficient $\beta_2$ naturally encodes information about the structure and dynamics of the near-horizon magnetosphere. In particular, its magnitude has been used to infer the relative importance of magnetic fields to the dynamics of the inflow (strong fields resist being wound into an otherwise turbulent plasma), and its phase has been shown to correlate with the ratio $B^\phi / B^r$ and to vary systematically with black hole spin (\citealt{palumbo_2020_beta2}; \citetalias{chael_2023_bhp1}). Ultimately, however, the details of the spin-dependence of the polarization arises from a combination of effects: the motion of the plasma, which sets the rotation rate of field lines, and general relativistic propagation effects near the horizon.

\subsection{Radial decomposition of \texorpdfstring{$\beta_2$}{beta2}}

\begin{figure*}[th!]
\centering
\includegraphics[width=1\linewidth]{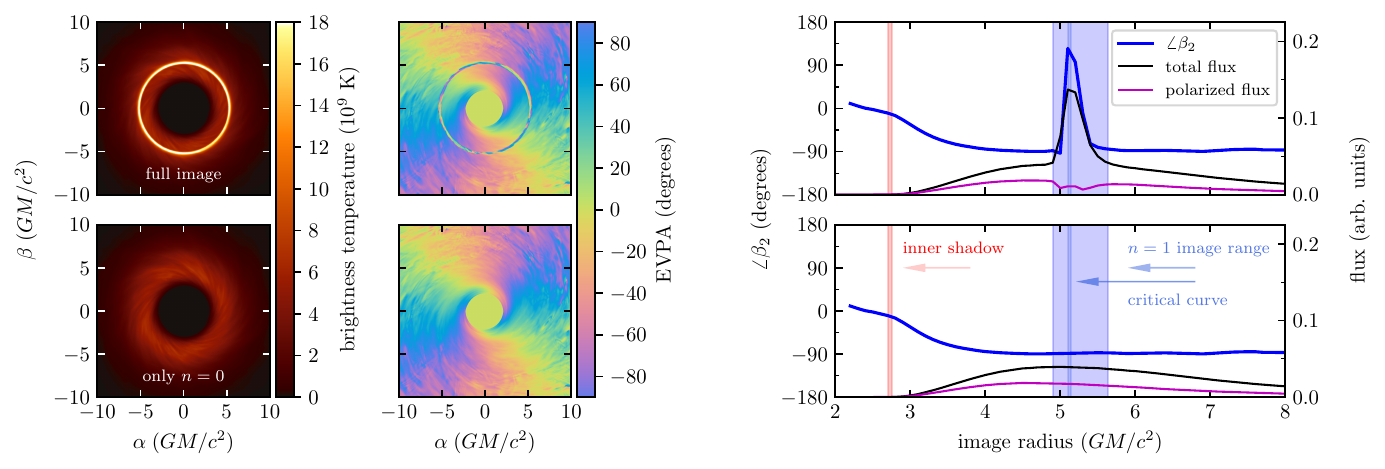}
\caption{Simulated black hole images and radial decomposition of the $\beta_2$ polarization mode. Left: Total intensity image. Center: Linear polarization map showing electric vector position angle (EVPA). Right: Radial profiles of Stokes I (black), polarized flux $\sqrt{Q^2 + U^2}$ (magenta), and the local $m=2$ polarization phase $\angle \beta_2(\rho)$ (blue). Top and bottom rows show the same image with and without contributions from $n > 0$ lensed emission. The image radii corresponding to the inner shadow (red) and the lensed image features (blue) are shown as colored bands.
The radial profile of $\angle \beta_2$ is determined by the magnetic field geometry, but is also influenced by boundary conditions imposed by the spacetime at the event horizon. In the direct emission region within the first photon subring, the phase varies smoothly and monotonically. The image-integrated value of $\angle \beta_2$ is not tied to a single radius, but reflects a flux-weighted average across annuli and can be dominated by regions of high polarized flux.
}
\label{fig:example_beta2_decomposed}
\end{figure*}

While the complex coefficient $\beta_2$ is often treated as a single, image-integrated quantity, evaluating it as a function of radius provides additional insight into how EVPA encodes the structure of the magnetosphere. By restricting the domain of the annulus in Equation~\ref{eqn:bzdefn}, it is possible to obtain a radial profile $\angle \beta_2(\rho)$, which reflects the azimuthal symmetry of the linear polarization at each radius. 

This annular decomposition is valuable because different radii in the image probe distinct photon trajectories and spacetime regions. Emission near the center of the image primarily originates from direct ($n = 0$) photon paths that graze the event horizon, where magnetic field lines are strongly anchored. In this region, $\angle \beta_2(\rho)$ exhibits a systematic spin dependence that arises from frame dragging and the growing azimuthal twist imposed on the magnetic field by black hole rotation. The radial variation and asymptotic behavior of $\angle \beta_2(\rho)$ as $\rho$ approaches the inner shadow encodes detailed information about how spin influences magnetospheric structure.

At larger radii, the image includes contributions from higher-order lensed photons, forming the photon ring. This ring introduces sharp features in $\angle \beta_2(\rho)$ that arise not from local field structure but from lensing geometry. While this component carries important information about spacetime, it complicates interpretation of the near-horizon magnetospheric physics. Isolating the direct-emission region therefore allows cleaner access to the field dynamics near the black hole.

Another key advantage of the radial $\angle \beta_2$ profile is its reduced sensitivity to uncertainties in the emissivity profile. Because the decomposition is performed annulus by annulus, it does not rely on assumptions about the global structure of the plasma or the relative strength of emission at different radii (quantities that are typically model-dependent and sensitive to the details of the accretion flow). Moreover, since the observed image-integrated $\angle \beta_2$ is effectively a flux-weighted sum of these local contributions, accurately reproducing the full radial structure provides a stringent test for any theoretical model.

Figure~\ref{fig:example_beta2_decomposed} shows a face-on polarimetric image from a GRMHD simulation of a black hole accretion flow in the MAD state, which is characterized by strong magnetic fields. Electron temperatures are set by the $T_{\rm ion}/T_{\rm electron}$ prescription of \citet{moscibrodzka_2016_rhigh}, with $R_{\rm low} = 1$ and $R_{\rm high} = 40$, which preferentially heats the ions in regions where the gas pressure far exceeds the magnetic pressure. The total intensity (left) displays the hot gas distribution, the shadow, and the photon ring, while the EVPA map (center) reveals a radially varying spiral pattern in the direct emission.\footnote{Neglecting the effects of Faraday rotation shifts the radially resolved values of $\angle \beta_2$ by $-5^\circ - -10^\circ$. The overall polarization pattern becomes more azimuthal, and the image-integrated value of $\beta_2$ becomes more negative by roughly $8^\circ$.}
This behavior is also clearly evident in the (right) azimuthally averaged radial profile of total intensity, linear polarization, and $\angle \beta_2$. At intermediate radii, the presence of the lensed photon ring introduces a sharp feature whose polarimetric signature is determined in large part by gravitational lensing and propagation effects rather than the local plasma parameters alone \citep[e.g.,][]{johnson_2020_universal,himwich_2020_universalpol,palumbo_2022_photonringbeta2}. The bottom row shows the same quantities as the top, but with contributions from the higher-order $n>0$ images removed. This isolates the smoother signal arising from direct emission, which also dominates the extended structure in the signal observed in ground-based EHT data.

The direct emission is more tightly coupled to the near-horizon magnetosphere physics and enables an easier comparison between different models.
Because the photon ring's contribution to $\angle \beta_2$ is well understood and largely geometric, we subtract it in what follows and focus on the direct $n = 0$ emission. This component carries the most information about the spin-dependent magnetic structure near the event horizon and forms the basis of our comparisons to GRMHD and semi-analytic models in the sections that follow.

\section{Near-Horizon Models of the Magnetosphere}
\label{sec:models}

We explore a spectrum of models for the near-horizon magnetosphere and use them to predict the polarization structure seen in black hole images. Comparing these models allows us to isolate the influence of individual physical ingredients and clarify how they shape the observed polarization patterns.

Our models span a range of physical complexity between the idealized, force-free split-monopole solution of \citet{blandford_1977_bz} to fully time-dependent numerical GRMHD simulations that self-consistently evolve the plasma and electromagnetic fields in the turbulent, inflowing-outflowing accretion system. Between these limits, we introduce a semi-analytic inflow model with a free magnetization parameter, which we show interpolates smoothly between force-free and GRMHD regimes. This model includes plasma inertia effects while remaining analytically tractable and computationally inexpensive.

Producing simulated images to compare to observations requires both the electromagnetic field, described by the Faraday tensor $F^{\mu\nu}$, and the plasma four-velocity, number density, and temperature, which set the synchrotron emission and absorption coefficients. We compute polarized images using the radiative transfer code \ipole \citep{moscibrodzka_2018_ipole} and the analytic ray tracing code {\tt kgeo} \citep{chael_2023_kgeo}. Our focus is on the resolved polarization pattern, i.e., the orientation of the EVPA across the image, which can be computed directly from the photon wavevector $k^\mu$ and $F^{\mu\nu}$. For this reason, we concentrate on differences in model prescriptions for $F^{\mu\nu}$, which can be expressed in terms of the fluid four-velocity $u^\mu$ and the magnetic field measured by an observer $B^i$ (\citetalias{chael_2023_bhp1}). We therefore focus on these quantities in the discussion below.

\subsection{Preliminaries}

We work in the Kerr metric, which is a stationary, axisymmetric vacuum solution to the Einstein field equations around an uncharged, rotating black hole. The Kerr metric has two parameters: the black hole mass $M$ and spin parameter $a \equiv J/M$, where $J$ is the angular momentum of the black hole. The outer event horizon in the Kerr spacetime is located at $r_+ = M + \sqrt{M^2 - a^2}$,
within which no causal signals can escape to infinity. The extremal limit corresponds to $a \to M$, where the horizon radius approaches $r_+ \to M$. In this work, we express the black hole angular momentum in terms of its dimensionless spin, $\bhspin \equiv a / M = J/M^2$, which normalizes its magnitude so that $\left| \bhspin \right| \le 1$. 
Hereafter, we work in geometric units ($G = M = c = 1$) and adopt Boyer-Lindquist coordinates  $(t, r, \theta, \phi)$. 

Under the assumption of stationarity and axisymmetry, the most general degenerate electromagnetic field can be written as the 2-form
{\small
\begin{align}
F = \dfrac{I(\psi)}{2\pi} \dfrac{\Sigma(r, \theta)}{\Delta(r) \sin\theta} \, \ud r \wedge \ud \theta
+ \ud \psi \wedge \left( \ud \phi - \Omega(\psi) \, \ud t \right),
\label{eqn:ansatz}
\end{align} }
where the field is fully determined by three functions: $\psi(r, \theta)$, $I(\psi)$, and $\Omega(\psi)$ (e.g., \citealt{gralla_2014_ffmagnetospheres}). This ansatz is applicable to both force-free electrodynamics and the electromagnetic sector of GRMHD. Physically, $\psi(r,\theta)$ encodes both the poloidal magnetic field geometry and the total magnetic flux enclosed by a surface of revolution around the spin axis. Its level sets correspond to field lines; in this way it can be viewed as a coordinate labeling poloidal magnetic field lines. The functions $I(\psi)$ and $\Omega(\psi)$ represent, respectively, the conserved electric current along a field line and the field line's angular velocity.

In the force-free limit, the field must satisfy $F_{\mu\nu} J^\nu = 0$, which reduces to a single nonlinear partial differential equation for $\psi(r,\theta)$, known as the Grad-Shafranov equation, where $I(\psi)$ and $\Omega(\psi)$ enter as free functions to be determined self-consistently. Solving this equation is highly nontrivial and generally requires numerical methods, though special solutions are known in simplified geometries. Although this formalism applies equally well to GRMHD fields, we make no assumptions about the governing equations of motion here and focus on the electromagnetic structure alone. The ansatz thus provides a unifying framework for comparing force-free, GRMHD, and semi-analytic inflow models.

To relate the electromagnetic field to observables, we compute the magnetic field in the frame of a fixed observer ``at rest at infinity.'' The magnetic field is then given by
\begin{align}
B^\mu = (\star F)^{\mu 0},
\end{align}
where $\hodgestar F$ is the Hodge dual of the field tensor, $(\star F)^{\mu\nu} = \frac{1}{2} \epsilon^{\mu\nu\alpha\beta} F_{\alpha\beta}$. The Levi-Civita tensor is $\epsilon^{\mu\nu\alpha\beta} = - [\mu\nu\alpha\beta]/\sqrt{-g}$, where $[\mu\nu\alpha\beta]$ is the antisymmetric symbol. This definition follows the conventional choices made for numerical GRMHD simulations (see, e.g., \citealt{gammie_2003_harm}). For the axisymmetric ansatz above (Equation~\ref{eqn:ansatz}), the magnetic field components reduce to
\begin{align}
\label{eqn:Br_defn}
B^r &= \dfrac{\partial_\theta \psi}{\Sigma(r,\theta) \sin\theta}, \\
\label{eqn:Bh_defn}
B^\theta &= -\dfrac{\partial_r \psi}{\Sigma(r,\theta) \sin\theta}, \\
\label{eqn:Bp_defn}
B^\phi &= \dfrac{I(\psi)}{2\pi \Delta(r) \sin^2\theta},
\end{align}
where $\Sigma(r,\theta) = r^2 + \bhspin^2 \cos^2\theta$ and $\Delta(r) = r^2 - 2 r + \bhspin^2$ are standard Kerr functions.

Since the polarization of synchrotron emission is oriented perpendicular to the magnetic field, it is only necessary to specify the relative ratios of the field components in order to study the polarimetric image structure. The field geometry, combined with the fluid velocity and photon wavevector, then determines the synthesized image. This motivates the use of reduced models, such as the inflow solution introduced below, which preserve the essential magnetic geometry while simplifying the plasma dynamics.

\subsection{Force-free split monopole solution}

To build intuition for the structure of the electromagnetic field near the black hole, we make use of the force-free split monopole solution of \citet{blandford_1977_bz}. This solution provides a simple, analytic model of an outgoing Poynting-dominated magnetosphere and serves as a valuable baseline for understanding more complex configurations. In the BZ solution, the electromagnetic field is described by a flux function $\psi(r, \theta)$, the field-line angular velocity $\Omega(\psi)$, and the current $I(\psi)$ consistent with the general stationary, axisymmetric, degenerate ansatz introduced above.

The stream equation derived from the force-free condition on the field ansatz of Equation~\ref{eqn:ansatz} is a nonlinear second-order PDE for the magnetic flux function $\psi(r,\theta)$, with additional dependence on the current $I(\psi)$ and field line rotation rate $\Omega(\psi)$. Solving this equation globally is analytically intractable, and even numerical approaches require significant care. However, the structure of the solution simplifies in the asymptotic limits as $r \to r_+$ near the horizon and as $r \to \infty$ at large distances. By leveraging these constraints, one can construct controlled approximations suitable for our analysis.

On the horizon, regularity requires the Znajek condition \citep{znajek_1977_condition}, which enforces that the flux, current, and field-line angular velocity combine so that $F^{\mu\nu}$ remains finite for infalling observers crossing $r = r_+$. On the other hand, spacetime becomes asymptotically flat in the limit $r \to \infty$, and to third order in spin, the electromagnetic field reduces to the well-known flat-space force-free monopole.

The typical treatment of the BZ split-monopole solution begins with the exact solution for a non-spinning black hole and then expands the exact perturbatively in orders of $\bhspin$, matching the Znajek condition order by order. In this paper, however, we introduce a simple approximation for the solution that agrees remarkably well and obeys the Znajek condition exactly on the horizon. More detail about the perturbative solution and the motivation for our approximate solution as well as a comparison between the two can be found in Appendix~\ref{app:bzmodel}. We express our model in terms of the ratio
\begin{align}
X(r,\theta) \equiv \dfrac{I}{2\pi(\Omega - \Omega_H) \, \partial_\theta \psi},
\end{align}
which can be used to compute the ratios between the $B^r, B^\theta$, and $B^\phi$ (Equations~\ref{eqn:Br_defn}--\ref{eqn:Bp_defn}). Our approximate solution naturally interpolates between the Znajek condition on the horizon and the monopole at infinity:
\begin{align}
\label{eqn:X1}
X_{\rm approx}(r,\theta) &= \frac{\sqrt{\Pi(r,\theta)} \, \sin\theta}{\Sigma(r,\theta)},
\end{align}
where
\begin{align}
\Pi(r,\theta) &= (r^2 + \bhspin^2)^2 - \bhspin^2 \Delta(r) \sin^2\theta.
\end{align}

The BZ split-monopole solution provides a physically transparent and analytically tractable model for black hole magnetospheres. Its key features include: a magnetic field structure anchored on the horizon and collimated along the rotation axis, a toroidal field component $B^\phi$ proportional to the conserved current $I(\psi)$, and field line angular velocity $\Omega(\psi) \sim \Omega_H / 2$, a requirement for extracting spin energy from the black hole. We use this solution as a baseline in what follows and examine to what extent its features persist in GRMHD simulations and in more general inflow models that incorporate finite fluid inertia.

\subsection{GRMHD overview}

To capture the dynamical structure of the black hole magnetosphere, which includes the behavior of both the magnetic field and the accreting plasma, we consider a set of time-dependent GRMHD simulations, which evolve a magnetized, relativistic fluid on a fixed Kerr background by solving a system of coupled conservation laws, under the assumption of ideal MHD (infinite conductivity and vanishing electric field in the fluid frame). We here provide a brief overview of the numerical simulation procedure. More detail about the simulation pipeline can be found in \citet{wong_2022_patoka}.

GRMHD codes evolve a set of eight independent variables, which can be expressed in terms of the rest-mass density $\rho$, internal energy $u$, pressure $P$, four-velocity $u^\mu$, and magnetic field four-vector $b^\mu = u_\nu \left(\star F\right)^{\nu\mu}$. The magnetic field in the coordinate frame is represented by the constrained variable $B^i \equiv (\star F)^{it} = b^i u^t - b^t u^i$, and the total stress-energy tensor includes both fluid and electromagnetic contributions,
\begin{align}
T^{\mu\nu} =& \left( \rho + u + P + b^{\lambda}b_{\lambda}\right) u^\mu u^\nu \\
& + \left(P + \frac{b^{\lambda}b_{\lambda}}{2} \right) g^{\mu\nu} - b^\mu b^\nu,
\end{align}
Written in a coordinate basis, the evolution equations are
\begin{align}
\partial_t \left( \sqrt{-g} \rho u^t \right) &= -\partial_i \left( \sqrt{-g} \rho u^i \right), \label{eqn:massConservation}\\
    \partial_t \left( \sqrt{-g} {T^t}_{\nu} \right) &= - \partial_i \left( \sqrt{-g} {T^i}_{\nu} \right) + \sqrt{-g} {T^{\kappa}}_{\lambda} {\Gamma^{\lambda}}_{\nu\kappa},  \\
\partial_t \left( \sqrt{-g} B^i \right) &= - \partial_j \left[ \sqrt{-g} \left( b^j u^i - b^i u^j \right) \right], \label{eqn:fluxConservation} \\
\partial_i \left( \sqrt{-g} B^i \right) &= 0, \label{eqn:monopoleConstraint} 
\end{align}
where $\sqrt{-g}$ is the determinant of the metric and ${\Gamma^\alpha}_{\beta\gamma}$ are Christoffel symbols and where the final expression imposes the no-monopole constraint on the magnetic field. The covariant ideal MHD condition $u_\mu F^{\mu\nu} = 0$ allows the electromagnetic field tensor $F^{\mu\nu}$ to be expressed entirely in terms of $u^\mu$ and $b^\mu$, closing the system.

Our simulations were performed with two GRMHD codes: \kharma \citep{prather_2024_kharma} and \koral \citep{sadowski_2013_koral}. Both codes solve the evolution equations in flux-conservative form on a Kerr background spacetime using high-resolution shock-capturing schemes. The \kharma simulations employed FMKS coordinates \citep{wong_2021_jetdisk} with a resolution of $384 \times 192 \times 192$ cells in the radial, latitudinal, and azimuthal directions and an outer boundary at $10^3\ GM/c^2$; further details are provided in \citet{wong_2022_patoka}. The \koral simulations used the cylindrified coordinates of \citet{ressler_2017_sgraelectrons} with a resolution of $288 \times 192 \times 144$ cells and an outer radial boundary fixed at $10^5 \, GM/c^2$; see \citet{narayan_2022_jetsurvey} for additional details. Both codes adopted an ideal gas equation of state with a uniform adiabatic index $\hat{\gamma} = 13/9$.

The set of GRMHD simulations we analyze corresponds to a subset of the fiducial libraries used in the EHT studies of M87$^*$ and \sgra \citep{eht_m87_5,eht_m87_8,eht_m87_9,eht_sgra_5,eht_sgra_8}, restricted to the magnetically arrested disk (MAD) accretion state. MAD models are characterized by their strong, ordered magnetic fields; our choice to restrict our attention to them is motivated both by their efficiency in powering BZ jets and by their consistency with EHT observations of M87$^*$ and \sgra.

In this work, we typically average our GRMHD simulation results over time and azimuth to produce a simple point of comparison against the other models. It is important to note, however, that the averaged simulation state is not necessarily a solution to the GRMHD equations, and an image produced from the averaged state is not equivalent to the average of images from the underlying simulation. Therefore there is no guarantee that the outcome of this averaging procedure will resemble a stationary, axisymmetric solution. Nevertheless, as we show in later sections, the averaged results provide a reasonably good basis for comparison.

\subsection{Axisymmetric equatorial inflow model}

GRMHD simulations self-consistently evolve both the fluid and electromagnetic fields but are computationally expensive. In contrast, force-free models capture the dynamics of highly magnetized plasmas without accounting for the fluid component, neglecting fluid inertia, velocity, and thermodynamics and assuming instead that the field dominates completely.

We develop a simplified, interpretable model that has a free parameter in the relative strength of the magnetic field compared to the fluid rest-mass energy density. Our approach solves a simplified version of the GRMHD equations under the ideal MHD approximation and features a free parameter related to the plasma magnetization, $\sigma \equiv b^2/\rho$, that governs the relative importance of electromagnetic and fluid inertia. In the high-$\sigma$ limit, the solution approaches force-free behavior; at moderate $\sigma$, inertial effects become significant, mimicking GRMHD. Our model is an extension of the inflow model proposed by \citet[][and see references within]{gammie_1999_inflow}.

We assume stationary, axisymmetric, equatorially symmetric, cold ($P = u = 0$) inflow in the Kerr spacetime, with dynamics confined to a thin wedge about the equatorial plane, and we specialize to prograde accretion. Under these conditions, the system reduces to a 1D radial problem with four conserved quantities
\begin{align}
\mathcal{F}_M &= 2\pi r^2 \rho u^r, \\
\mathcal{F}_L &= 2\pi r^2 T^r{}_\phi, \\
\mathcal{F}_E &= -2\pi r^2 T^r{}_t, \\
\Phi_B &= \pi F_{\theta\phi},
\end{align}
representing mass, angular momentum, energy, and magnetic fluxes, respectively.

The model is specified by: 
\begin{itemize}
\item the black hole spin,
\item the radial magnetic field $F_{\theta\phi}$, which we express in terms of the conventional MAD parameter $\tilde{\phi} \sim F_{\theta\phi} \sqrt{\pi} / 2$ \citep{tchekhovskoy_2011_mad,porth_2019_sanecomp},\footnote{We find this relation under the assumption that accretion is confined to a uniform wedge about the midplane. We write $\tilde{\phi}$ rather than $\phi$ to acknowledge the geometric differences between the two formulations. Our paper uses Lorentz-Heaviside units such that the MAD saturation value is $\phi \approx 15$.}
\item the location of the outer radial boundary where $u^r = 0$, and
\item and the angular velocity of the fluid $\Omega_{\rm Fluid} = u^\phi/u^t$ at the boundary. 
\end{itemize}
Notice that the stationary axisymmetric equations require $\Omega_{\rm Field} = \Omega_{\rm Fluid}$ where $u^r = 0$, so choosing the location of the boundary and $\Omega_{\rm Fluid}$ at that point also determines the field line rotation rate, which is constant across radius.

The system is closed by the requirement that the solution pass smoothly through the fast magnetosonic point, where the flow speed matches the characteristic wave speed. This regularity condition selects a unique value of the angular momentum flux $\mathcal{F}_L$, which completes the system and determines the radial structure of the solution. The resulting model is computationally efficient, requiring only root-finding and algebraic evaluation rather than time integration. More information about the implementation details are provided in Appendix~\ref{app:inflow_model}.

\section{The Structure of the Emission Region}
\label{sec:model_comparison}

We begin by examining the structure of the inflow solutions in terms of the underlying fluid and field variables and then assess how well these solutions reproduce the qualitative trends observed in GRMHD simulations. This comparison allows us to identify the aspects of the dynamics that are robust to model assumptions as well as the places where the model breaks down.
We thus probe the validity of the inflow model as a framework for exploring controlled departures from the force-free limit and for efficiently probing parameter space around GRMHD simulations.

\begin{figure*}[htp!]
\centering
\includegraphics[width=0.85\linewidth]{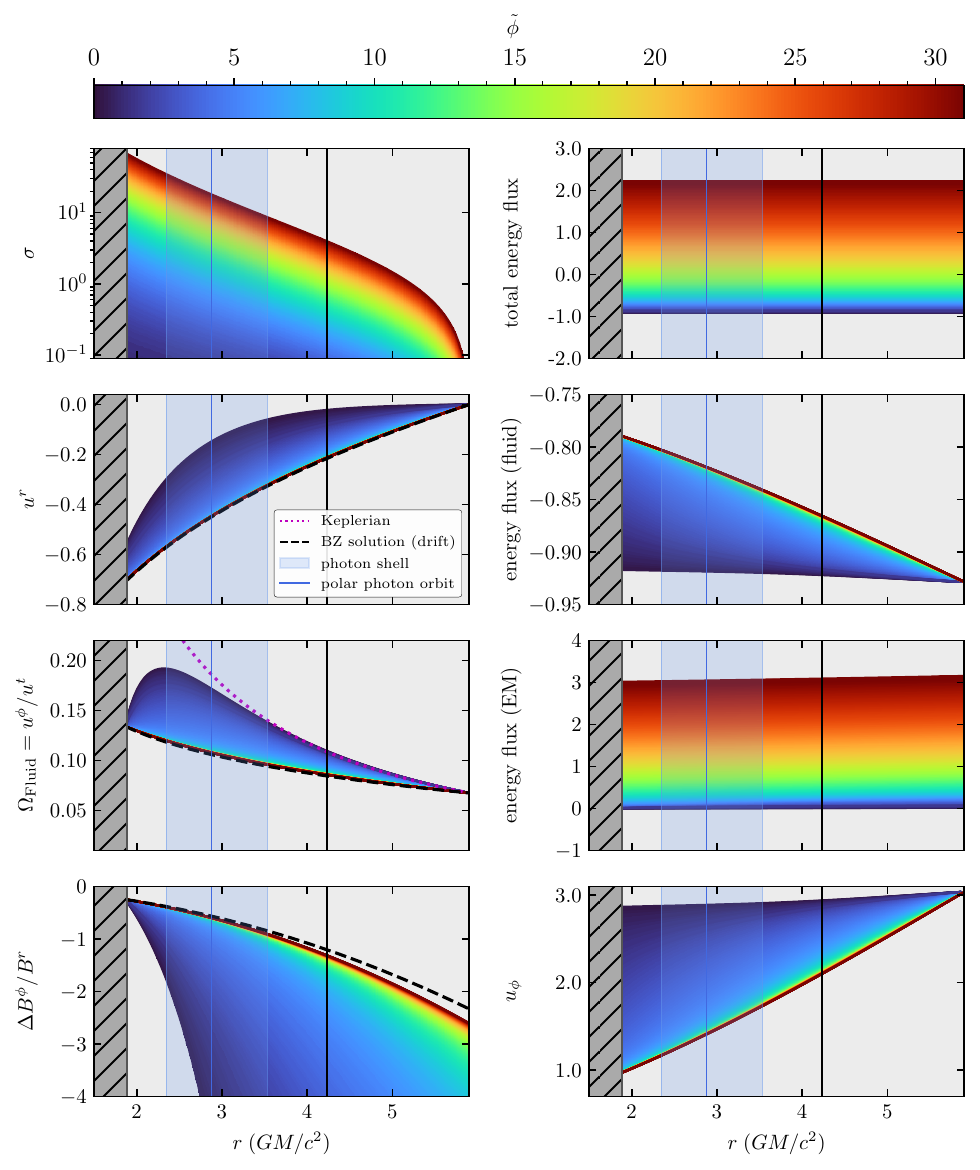}
\caption{
Midplane inflow solutions for a black hole with spin $\bhspin = 0.5$ and fixed outer boundary $r_{\rm bdd} \approx 6$, where $\Omega_{\rm Fluid} = \Omega_{\rm Field} \approx \Omega_{\rm BZ} \approx 0.06725$. Each curve corresponds to a different value of magnetic flux $\tilde{\phi}$ ranging from $0$ to $31$. Vertical shaded regions mark the photon shell (light blue) and polar photon orbit (dark blue), which is visible to face-on observers. The vertical black line marks the location of the innermost stable circular orbit.
Dashed black lines show the approximate force-free split-monopole solution for $u^r$, $\Omega_{\rm Fluid}$, and the magnetic field winding $\Delta B^\phi / B^r$, with four-velocities set to the unique drift velocity that conserves energy for a cold flow. The dotted purple curve in the $\Omega_{\rm Fluid}$ panel shows the Keplerian angular velocity for comparison.
As $\tilde{\phi}$ increases, the flow becomes more magnetically dominated and increasingly force-free. With increased $\sigma$, the radial infall velocity $u^r$ steepens, the angular velocity $\Omega_{\rm Fluid}$ drops, and magnetic winding $\Delta B^\phi/B^r$ decreases to approach the split-monopole prediction. Electromagnetic energy flux increases significantly with $\tilde{\phi}$, while the fluid contribution to the energy flux remains nearly constant.
}
\label{fig:inflow_model}
\end{figure*}

\subsection{The dependence on the magnetization}

We begin by examining how the inflow solution depends on the magnetic flux $\tilde{\phi}$ while holding other parameters fixed. Our fiducial case considers a black hole with spin $\bhspin = 0.5$ with the outer boundary placed at $r_{\rm bdd} \approx 6\ GM/c^2$. This choice sets the boundary relatively far from the horizon while still allowing for a field line rotation rate close to the BZ value, $\Omega_{\rm Field} \approx \Omega_{\rm BZ} \approx 0.06725$, which will allow for a field structure that matches the BZ monopole as closely as possible. We impose $u^r = 0$ at this radius and enforce corotation between field and plasma.

\begin{figure*}[htp!]
\centering
\includegraphics[width=\textwidth]{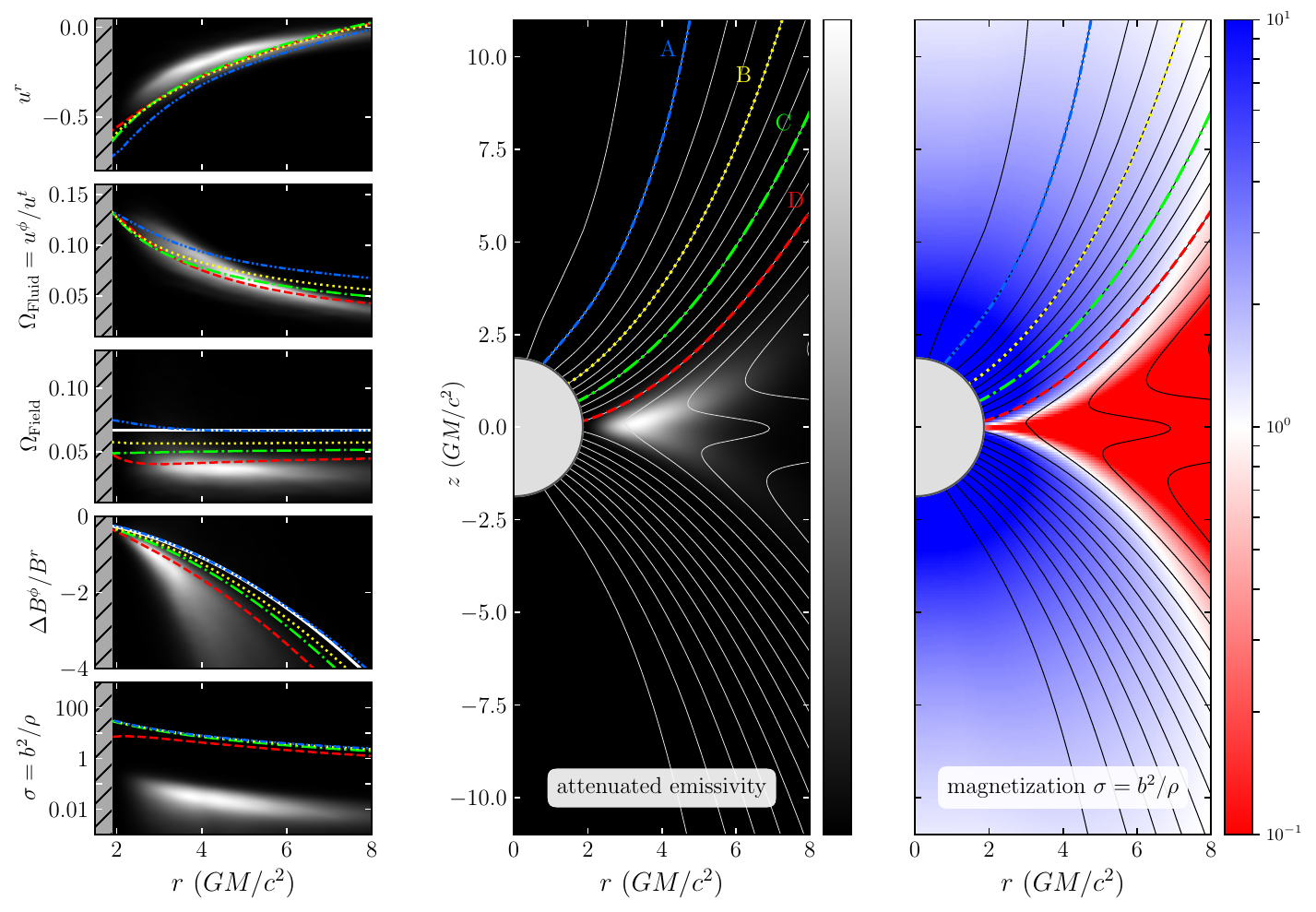}
\caption{
Location and properties of the emission region in a time- and azimuthally averaged GRMHD simulation of a MAD accretion flow with black hole spin $\bhspin = 0.5$. The grey-scale maps in the left and center panels show the distribution of the attenuated emissivity (i.e., accounting for absorption along the line of sight to an observer at inclination $163^\circ$) as a function of radius and local fluid properties. In the $\Omega_{\rm Field}$ and $\Delta B^\phi/B^r$ panels, the white dashed lines indicate the corresponding profiles from the force-free Blandford-Znajek (BZ) solution, which has $\Omega_{\rm Field} \approx 0.06725$.
The right panel shows the magnetization in the same averaged GRMHD simulation; regions with higher $\sigma$ are closer to force free. The dashed, colored lines in each panel denote magnetic field lines chosen at increasing latitudes angles (labeled A--D, with $A$ at the higher latitude). The profile of the same fluid quantities along each line are shown in the left panels.
Although most of the observed emission originates from regions with modest magnetization, the solution kinematics and magnetic field structure remain broadly similar to that in the highly magnetized regions. This supports the use of idealized models in interpreting the global field geometry even when emission is produced in regions with finite $\sigma$.
}
\label{fig:grmhd_emsource_comparison}
\end{figure*}

Figure \ref{fig:inflow_model} shows the resulting family of inflow solutions across a range of $\tilde{\phi}$. The shaded gray region marks the event horizon; the vertical blue band highlights the photon shell, and the dark blue line within it identifies the polar photon orbit, which produces the critical curve for a face-on observer. The innermost stable circular orbit (ISCO) is shown as a solid black line. For comparison against the fluid velocity, we plot the Keplerian rotation rate as a dotted purple curve. We also show traces from the approximate split-monopole force-free solution for the magnetosphere in the panels for $u^r, \Omega_{\rm Fluid}$, and $\Delta B^\phi/B^r$ (dashed black; hereafter, we multiply the ratio by $\Delta(r)$ to avoid the coordinate singularity at the event horizon). The force-free velocities are computed from the field configuration by selecting the drift velocity and choosing the unique (radius-dependent) boost parallel to the magnetic field that ensures conservation of energy (see \S3.4 of \citealt{gelles_2025_spinjetpol}).

As $\tilde{\phi}$ increases, the inflow becomes progressively more magnetically dominated, with a corresponding rise in magnetization $\sigma$. Approaching the force-free regime yields several systematic trends:
\begin{itemize}[leftmargin=2em]
\item The infall velocity $u^r$ steepens across models and its slope tends toward a constant value, indicating that there is less of the rapid acceleration characteristic of plunging flows.
\item The fluid angular velocity $\Omega_{\rm Fluid} = u^\phi/u^t$ decreases, reflecting reduced angular momentum per baryon due to electromagnetic stresses more efficiently removing angular momentum from the fluid.
\item The degree of field winding, quantified by $|\Delta B^\phi / B^r|$, decreases and approaches the force-free split-monopole value.
\item The electromagnetic contribution to the total energy flux grows steadily, while the fluid contribution remains nearly constant.
\end{itemize}

The parameter $\tilde{\phi}$ evidently determines the balance between plasma inertia and magnetic dominance in the inflow. At low values, fluid inertia plays a significant role in modifying the rotation profile and enhancing field winding through the plasma’s own rotation. At high values, the magnetosphere field governs the dynamics, and the solution converges toward the force-free limit. Varying $\tilde{\phi}$ thus provides a way to continuously transition between the inertia-dominated and magnetically dominated regimes, allowing for controlled exploration of the effects of the fluid on the large-scale structure of the magnetosphere.

\subsection{Comparison to GRMHD simulations}

Having established how the inflow solution depends on magnetic flux, we next assess how well the model reproduces the trends found in full numerical simulations. We compare the predictions of the inflow model against a time- and azimuthally averaged GRMHD simulation of a magnetically arrested disk (MAD) around a black hole with spin $\bhspin = 0.5$; this is the same simulation used for the synthetic image analysis in Figure~\ref{fig:example_beta2_decomposed}.

Figure~\ref{fig:grmhd_emsource_comparison} summarizes the properties of the GRMHD simulation within its effective emission region, defined by how much the local emissivity contributes to the observed image.\footnote{To compute this attenuated emissivity, we integrate the optical depth $\tau$ along photon geodesics traced backward from the observer through the simulation domain. At each point along the geodesic, the local emissivity is multiplied by the cumulative transmission factor $e^{-\tau}$, which accounts for absorption and obscuration along the line of sight. The resulting attenuated contributions are then accumulated over the coordinates $(r,\theta)$ to produce a two-dimensional spatial map of where the observed emission originates within the flow.} Our calculation accounts for optical depth effects, but they only decrease the total flux received by $\approx 9\%$, and neglecting the importance of optical depth does not qualitatively change the structure of the emission region.

The grey-scale maps in left and center panels of Figure~\ref{fig:grmhd_emsource_comparison} show the distribution of attenuated emissivity as a function of radius and several physical quantities that shape the observed polarimetric structure (the white regions are where most of the emission originates): the fluid radial velocity $u^r$, the fluid angular velocity $\Omega_{\rm Fluid}$, the field line rotation rate $\Omega_{\rm Field}$ (computed following Equation D148b of~\citetalias{chael_2023_bhp1}), and the degree of magnetic field winding $\Delta B^\phi/B^r$. We also plot the magnetization $\sigma$ in the emission regions, which allows us to gauge how close they are to force-free.

The bottom-left, center, and right panels of Figure~\ref{fig:grmhd_emsource_comparison} show that much of the observed emission originates below the funnel wall within the upper layers of the accretion flow. These regions are only modestly magnetized ($\sigma \lesssim 1$) and therefore occupy an intermediate regime in which both plasma inertia and magnetic stresses play important roles in shaping the dynamics. Even outside the force-free limit, however, the field structure in these regions exhibits clear radial trends that remain strikingly consistent with the force-free solution and the inflow model.

Each panel also shows profiles of the same quantities evaluated along four representative magnetic field lines (colored lines; A-D), selected at progressively higher polar angles where the flow is closer to force-free. As the field lines approach higher latitudes, their behavior increasingly resembles the force-free BZ solution. For comparison, the expected BZ trends are shown as white dashed curves in the $\Omega_{\rm Field}$ and $\Delta B^\phi/B^r$ panels.

In contrast, the emission-weighted regions show systematically lower field-line rotation rates, more tightly wound magnetic fields, and fluid angular velocities $\Omega_{\rm Fluid}$ that fall below the Keplerian profile. This departure from the idealized BZ solution highlights the role of plasma inertia in the emission zone and supports the interpretation that the observed radiation originates in a regime not captured by force-free models alone. Nevertheless, the underlying field structure remains coherent and approximately monopolar in topology, suggesting that the inflow model provides a reasonable approximation to the salient features of the time-dependent GRMHD solution.

\vspace{0.5em}

\begin{figure}[htp!]
\centering
\includegraphics[width=\linewidth]{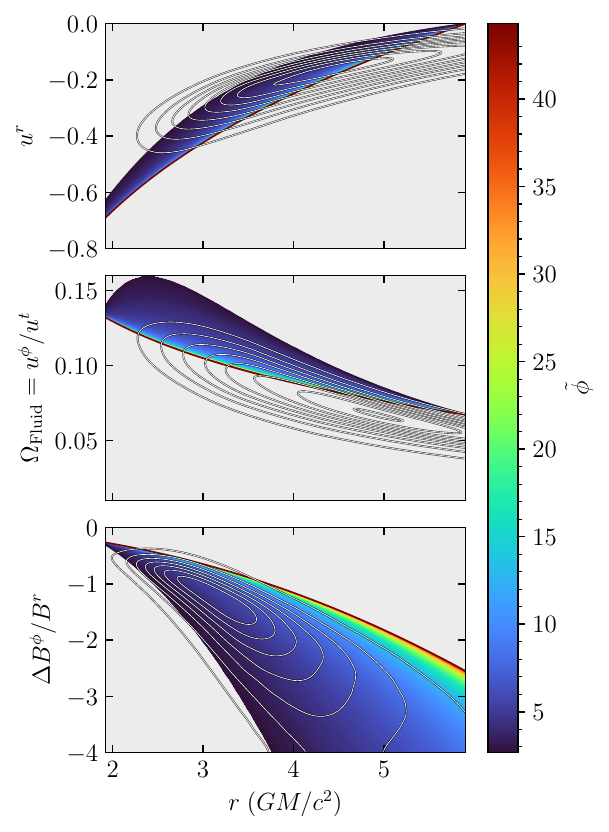}
\caption{
Comparison between inflow model and GRMHD simulation with $\bhspin = 0.5$ and inflow model outer boundary $r_{\rm bdd} \approx 6M$.
Each panel shows inflow model predictions for (top) radial velocity $u^r$, (center) fluid angular velocity $\Omega_{\rm Fluid} = u^\phi/u^t$, and (bottom) magnetic field winding $|\Delta B^\phi / B^r|$ as functions of radius and magnetic flux $\tilde{\phi}$. Color encodes the inflow model prediction. Overlaid contours show the GRMHD values of the same quantities at each radius, weighted by the attenuated emissivity, and are plotted at levels $0.1, 0.2, \ldots, 0.9$.
Across these parameters, the inflow model recovers the general behavior of these quantities in the simulation, although it differs quantitatively in the fluid velocity.
}
\label{fig:model_grmhd_comparison_grmhd_Omega}
\end{figure}

How well do the solutions of the inflow model reproduce the behavior seen in GRMHD? Figure~\ref{fig:model_grmhd_comparison_grmhd_Omega} compares the inflow solutions of Figure~\ref{fig:inflow_model} against the simulation for the three quantities most directly tied to the polarimetric structure of the image: the radial velocity $u^r$, the fluid-frame angular velocity $\Omega_{\rm Fluid}$, and the magnetic winding $\Delta B^\phi / B^r$. For reference, we overlay contours of the emission-weighted profiles derived from the same GRMHD simulation used in Figures~\ref{fig:example_beta2_decomposed} and \ref{fig:grmhd_emsource_comparison}, using the attenuated emissivity as a proxy for the regions that contribute the observed image.

Across all three quantities, the inflow model reproduces the qualitative radial trends seen in the GRMHD results. As $\tilde{\phi}$ increases, i.e., as the effects of the fluid inertia are decreased, the solutions show weaker field winding and lower $\Omega_{\rm Fluid}$. These patterns align with the structure seen in the GRMHD simulation, since the emissivity-weighted regions align most closely with inflow solutions of moderate magnetization. In particular, a narrow range of values, $\tilde{\phi} \approx 5 - 7$, produces inflow solutions that track the GRMHD contours across a broad range of radii. Although this agreement is somewhat surprising given that the most magnetized regions in the simulations are not equatorial, the comparison suggests that the inflow model captures underlying features of the dynamics that are insensitive to the precise geometry of the magnetosphere.

Despite the overall agreement, some discrepancies remain. One of the clearest is in the behavior of the radial infall velocity, with the inflow model predicting systematically faster accretion than is found in the GRMHD simulation. This difference reflects the model’s cold-plasma assumption, which omits thermal pressure gradients that act to slow the infall in hotter flows. Yet despite the deviations, the inflow model still captures the leading-order dynamics and remains useful as an interpretive tool.

\begin{figure*}[htp!]
\centering
\includegraphics[width=\linewidth]{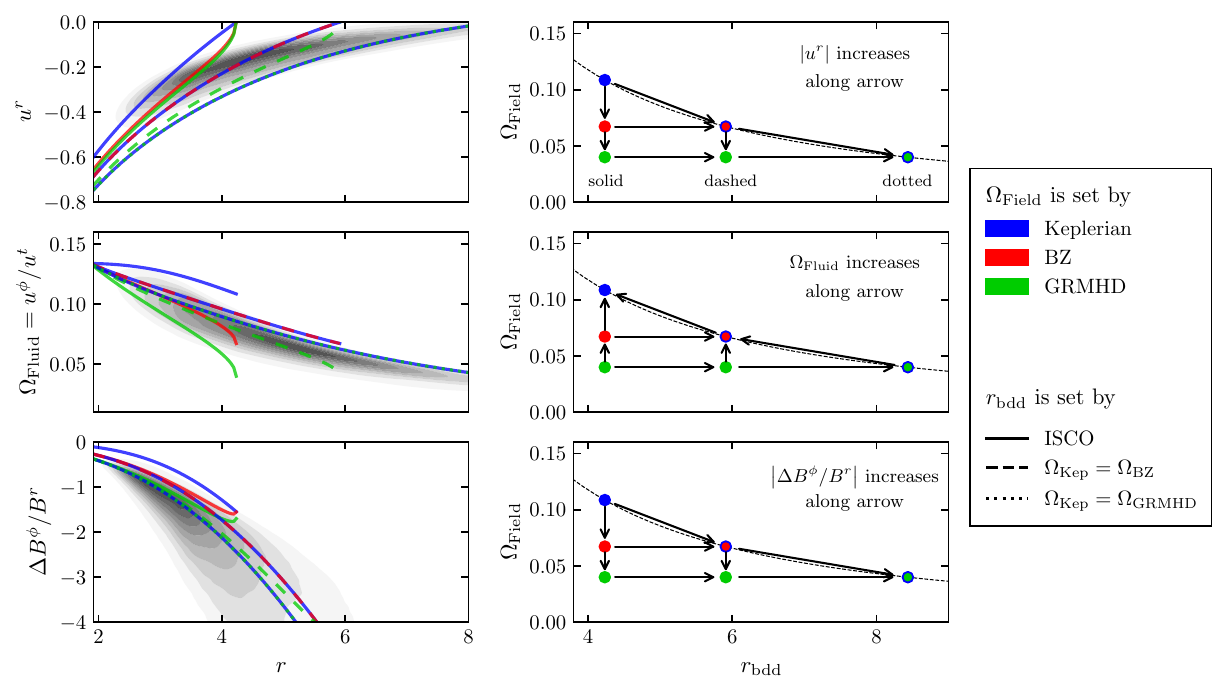}
\caption{
Comparison of inflow solutions for fixed black hole spin ($\bhspin = 0.5$) and magnetic flux ($\tilde{\phi} \approx 7$), across a range of outer boundary conditions. The left panels show the radial profiles of $u^r$, $\Omega_{\rm Fluid} = u^\phi/u^t$, and $\Delta B^\phi/B^r$ for different boundary radii $r_{\rm bdd}$ and field line angular velocities $\Omega_{\rm Field}$. Background shading shows the corresponding GRMHD values, weighted by attenuated emissivity, with contours at $0.1, 0.2, \ldots 0.9$ of the maximum weighting. The right panels summarize how changing $r_{\rm bdd}$ and $\Omega_{\rm Field}$ impacts the behavior of the radial profiles. Each point in the right panel represents a full solution (radial profile) in the corresponding left panel. The regions in the upper-right portion of parameter space are excluded as unphysical, since they correspond to fluid rotation that is too rapid to allow axisymmetric inflow.
Larger $r_{\rm bdd}$ or smaller $\Omega_{\rm Field}$ increases $|u^r|$ and the field winding $\Delta B^\phi/B^r$. $\Omega_{\rm Fluid}$ increases with $\Omega_{\rm Field}$, but also with $r_{\rm bdd}$. The cold inflow model captures key GRMHD trends but cannot fully match the GRMHD-preferred region, since decreasing the radial fluid velocity also decreases $\Delta B^\phi/B^r$. Inclusion of gas pressure (neglected here) could slow the flow while maintaining similar magnetic structure.
}
\label{fig:inflow_model_boundary_conditions}
\end{figure*}

\subsection{Boundary conditions for the inflow model}

The boundary conditions adopted in the inflow solutions of Figure~\ref{fig:inflow_model} do not perfectly match the fluid variables measured in the GRMHD simulation. This naturally raises the question of whether the discrepancy in velocity profiles could be alleviated by adjusting the boundary prescription, e.g., by moving the outer boundary of the inflow model to larger radii. Unfortunately, such a modification introduces two complications. First, because the model enforces $u^r = 0$ and assumes circular fluid motion at the boundary, the fluid angular velocity cannot be increased arbitrarily at large radius without invoking additional inward forces. Second, shifting the boundary outward inevitably reduces the fluid’s rotation rate and, in turn, the rotation rate of the field lines. This limitation is not necessarily a flaw, since our goal is not to reproduce the Blandford–Znajek solution in exact detail but to systematically explore deviations from it. Since the inflow framework is designed to allow for variation in the field-line rotation rate, it is natural to explore solutions beyond the BZ limit in this parameter as well.

We thus test whether varying the boundary parameters can bring the inflow model into closer agreement with the GRMHD simulation. Figure~\ref{fig:inflow_model_boundary_conditions} shows how the solutions respond to changes in $r_{\rm bdd}$ and $\Omega_{\rm Field}$ at fixed black hole spin $\bhspin = 0.5$ and magnetization $\tilde{\phi} \approx 7$. We find solutions for three boundary locations: the innermost stable circular orbit at $r\approx 4.23\ GM/c^2$ and the two radii where the Keplerian angular velocity matches either $\Omega_{\rm BZ}$ or the GRMHD-inferred value $\Omega_{\rm GRMHD} \approx 0.04$ ($r_{\rm bdd} \approx 5.91$ or $8.44\ GM/c^2$, respectively). In the left column of Figure~\ref{fig:inflow_model_boundary_conditions}, each colored curve shows a radial profile from one inflow solution overlaid on the emission-weighted GRMHD contours. The right column summarizes the range of inflow solution values as a function of radius for each choice of boundary conditions. The regions in the upper-right portion of parameter space are excluded as unphysical, since they correspond to fluid rotation that is too rapid to allow axisymmetric inflow.

Increasing the outer boundary radius $r_{\rm bdd}$ and lowering the field-line rotation rate both steepen the radial infall: with a larger acceleration zone, the plasma gains more inward momentum, and slower rotation reduces centrifugal support and drives the flow closer to freefall. These same changes also enhance magnetic field winding. As a result, no single set of boundary conditions allows the inflow model to reproduce all of the features of the GRMHD simulation simultaneously. Boundary choices that match the degree of magnetic winding $\Delta B^\phi / B^r$ tend to overestimate the radial velocity $u^r$, and boundary conditions that suppress $|u^r|$ produce under-wound fields. This tension again reflects the limitations of the cold inflow framework: without pressure support, the plasma cannot maintain slower infall or more gradual rotation without plunging into the black hole. Even so, the inflow solutions still reproduce the qualitative structure of both the fields and the flow.

\section{Polarimetric Signatures of the Inflow Model}
\label{sec:polarization_spin}

\begin{figure*}[htp!]
\centering
\includegraphics[width=\linewidth]{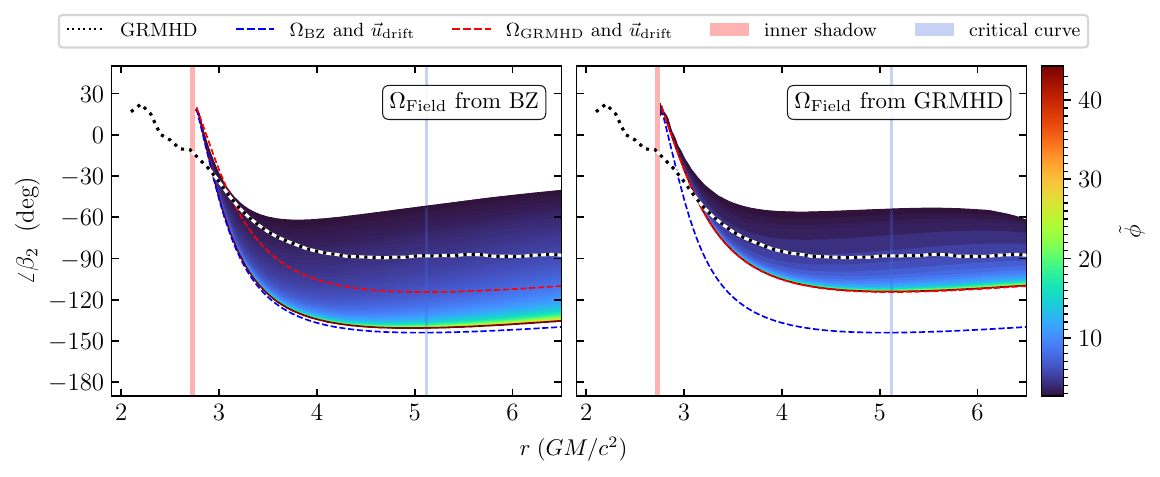}
\caption{
Comparison of radial profiles of $\angle \beta_2$ between inflow models and GRMHD simulations for $a_\star = 0.5$.
Colored lines show inflow model predictions for a nearly face-on observer ($i = 179^\circ$), while the thick black curve shows the averaged result from a GRMHD simulation. The red shaded region marks the location of the inner shadow \citep{chael_2021_innershadow}, and the blue band indicates the critical curve (asymptotic photon ring).
The left and right columns correspond to inflow solutions with $\Omega_{\rm Field} = \Omega_{\rm BZ}$ and $\Omega_{\rm Field} = \Omega_{\rm GRMHD}$, respectively. As $\tilde{\phi}$ increases, the solutions transition from inertia-dominated to more force-free behavior, driving $\angle \beta_2$ toward the limiting profile of a split-monopole magnetic field with the specified rotation rate (and velocity set by the corresponding drift velocity). At low $\tilde{\phi}$, the field becomes more tightly wound, shifting $\angle \beta_2$ closer to radial. The variation of the inflow solution across values of $\tilde{\phi}$ illustrates how the model smoothly interpolates between force-free and inertia-dominated regimes. The agreement between the inflow model and the emissivity-weighted GRMHD behavior follows the same trends as the best-fit $\tilde{\phi}$ values in Figure~\ref{fig:model_grmhd_comparison_grmhd_Omega}. This consistency supports choosing $\tilde{\phi} \simeq 5-7$ as a good proxy for reproducing GRMHD simulations. Note that the $\angle \beta_2$ curve obtained from the simulation is faint and jet-dominated near and within the inner shadow, and consequently should not be expected to agree with the model.
}
\label{fig:beta2_vs_radius}
\end{figure*}

We now look at the polarimetric predictions of the inflow model, beginning with the radial dependence of $\angle \beta_2$ and then considering the image-integrated value. We compare the inflow model predictions to those obtained from time-dependent GRMHD simulations to guide our understanding of the relationship between the polarization structure observed in simulated images and the parameters of our inflow model. Finally, we use the model to demonstrate how $\angle \beta_2$ can serve as a diagnostic of black hole spin, showing that for realistic model parameters the observable distribution of $\angle \beta_2$ provides a means to distinguish between different values of spin.

\subsection{Radial profiles of \texorpdfstring{$\angle \beta_2$}{arg(beta2)}}

We begin by computing the radial profiles of $\angle \beta_2$ predicted by the inflow model for a range of values of $\tilde{\phi}$ under the two representative boundary conditions in Figure~\ref{fig:inflow_model_boundary_conditions}. In both cases the boundary radius is fixed at $r_{\rm bdd} \approx 6\ GM/c^2$, but one adopts $\Omega_{\rm Field} = \Omega_{\rm BZ}$ while the other uses the lower field line rotation rate $\Omega_{\rm Field} \approx \Omega_{\rm GRMHD}$. These choices are motivated by the fact that they produce fluid profiles most consistent with those recovered from emissivity-weighted GRMHD simulations and thus provide a physically motivated basis for comparison. 

Figure~\ref{fig:beta2_vs_radius} shows the radial profiles of $\angle \beta_2$ predicted by the inflow model across a range of $\tilde{\phi}$ values compared directly to the GRMHD simulation and to the force-free split-monopole limit (with the fluid motion in the force-free model set to the drift velocity). The inflow solutions recover the same qualitative trends as the simulation traces of $\angle \beta_2$ for both choices of boundary condition. At low $\tilde{\phi}$, when fluid inertia dominates, the field lines are wound more tightly, and as $\tilde{\phi}$ increases, the influence of inertia weakens, and the inflow profiles converge smoothly toward the force-free limit. Values of $\tilde{\phi}\sim5-7$ produce profiles in close quantitative agreement with the GRMHD results across most radii, with the largest deviations at small radii where off-midplane emission (e.g., from the jet funnel) contaminates the midplane inflow prediction. Unsurprisingly, agreement improves when $\Omega_{\rm Field}\approx \Omega_{\rm GRMHD}$, which shows that both the rotation profile and the magnetization $\tilde{\phi}$ are critical in determining the synchrotron polarization.

\vspace{0.5em}

How effective is polarization as a diagnostic of black hole spin? To address this question, we begin by examining the radial profiles of $\angle \beta_2$ predicted by the inflow model across different spins. To maintain consistency with numerical simulations, we compute the emissivity-weighted average field-line rotation rate in a suite of MAD simulations performed with the \kharma code. The resulting dependence is well described by a simple linear fit,
\begin{align}
\Omega_{\rm Field} \approx 0.05\,\bhspin + 0.015,
\label{eqn:omega_field_fit}
\end{align}
which reflects both the nonzero field line rotation due to the angular momentum of the inflowing plasma even at $\bhspin = 0$ as well as the monotonic increase with spin driven by frame-dragging of the black hole magnetosphere.

\begin{figure}[htp!]
\centering
\includegraphics[height=20em]{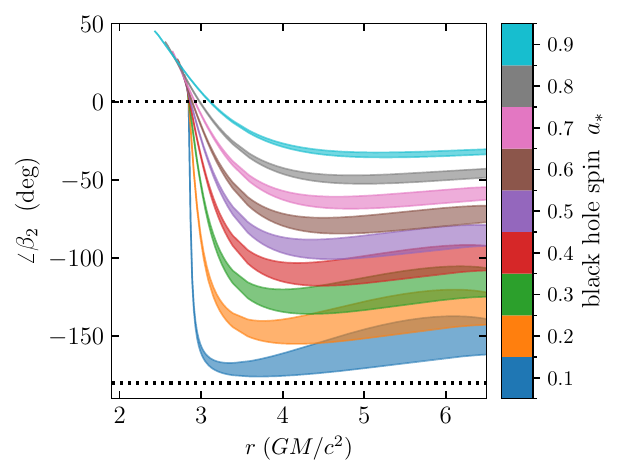}
\caption{
Radial profiles of $\angle \beta_2$ computed from the inflow midplane model, viewed at low inclination, for $\tilde{\phi} \in [5, 7]$ and $\Omega_{\rm Field}$ set by the fitting function of Equation~\ref{eqn:omega_field_fit}. Each colored band corresponds to a different black hole spin. For all spins, $\angle \beta_2$ rapidly approaches a characteristic asymptotic value near the edge of the inner shadow, consistent with the universal limit and with small differences reflecting finite numerical resolution.
Assuming negligible (or correctable) Faraday rotation, the image-integrated values of $\angle \beta_2$ predicted by the model remain clearly distinguishable between spins for these intermediate values of $\tilde{\phi}$ (chosen to be consistent with the $\bhspin = 0.5$ simulation).
}
\label{fig:spin_library_comparison}
\end{figure}

Figure~\ref{fig:spin_library_comparison} shows the radial profiles of $\angle \beta_2$ predicted by the inflow model for nine black hole spins spanning $\bhspin = 0.1-0.9$. For each spin, we vary $\tilde{\phi}$ over the range $5-7$ and set $\Omega_{\rm Field}$ according to Equation~\ref{eqn:omega_field_fit}. In all cases, $\angle \beta_2$ converges to its spin-dependent asymptotic value at the edge of the inner shadow. At larger image radii, however, the profiles remain clearly separated by spin. The systematic dependence on $\bhspin$ found in GRMHD simulations can therefore be understood as the outcome of the competition between force-free rotation, set by the horizon and frame-dragging, and rotational motion induced by plasma inertia. The inflow model provides a transparent way to disentangle these effects: $\Omega_{\rm Field}$ controls the baseline rotation inherited from the black hole, while $\tilde{\phi}$ tunes how much the plasma inertia can alter the field and shift $\angle \beta_2$ away from the split-monopole limit.

\begin{figure*}[htp!]
\centering
\includegraphics[height=20em]{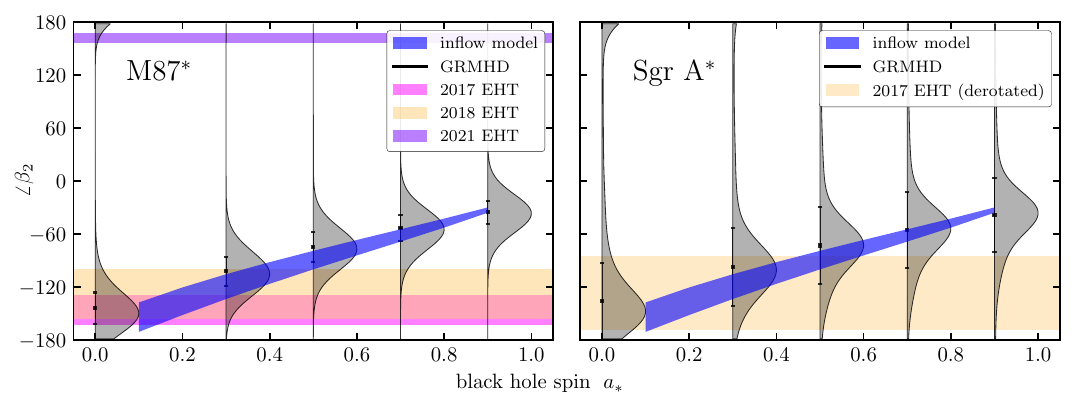}
\caption{
Comparison of image-averaged values of $\angle \beta_2$ as a function of black hole spin, computed from either the midplane inflow model (blue band) or GRMHD simulations of MAD accretion flows (gray band). Data and simulations for the M87$^*$ accretion system are shown on the left and for the \sgra system are shown on the right. Simulations for \sgra are over inclinations $i=10, 30, 50, 70^\circ$ relative to the spin axis of the black hole, which is directed away from us. The inflow model uses a range of magnetizations ($5 \le \tilde{\phi} \le 7$) and field line angular velocities $\Omega_{\rm Field}$ inferred from GRMHD data.
All models show the same qualitative trend: increasing spin leads to more tightly wound magnetic fields and a more radial polarization pattern. The inflow model closely tracks the GRMHD simulations, which supports the interpretation of $\angle \beta_2$ as a geometric tracer of spin and magnetospheric structure in the presence of fluid inertia.
The observed ranges of $\angle \beta_2$ for M87$^*$ for different years are shown as a horizontal magenta, orange, and purple bands, and the range for \sgra is shown as a horizontal orange band. Robustly inferring the spin from observations will require correcting for Faraday rotation, which may be achievable with multifrequency measurements.
}
\label{fig:spin_grmhd_inflow_comparison}
\end{figure*}

Although the model presented here provides a compelling explanation of the qualitative relationship between black hole spin and the observed polarimetric structure, the quantitative relationship between spin and polarization depends on selecting values both for the field line rotation rate $\Omega_{\rm Field}$ as well as the magnetization of the plasma $\tilde{\phi}$. Indeed, Figure~\ref{fig:beta2_vs_radius} reveals a strong dependence of $\angle \beta_2$ on the value of $\tilde{\phi}$, with smaller values of $\tilde{\phi}$ producing more radial polarization patterns. The bands in  Figure~\ref{fig:spin_library_comparison} show $\angle\beta_2$ calibrated to GRMHD MAD simulations.   However, the level of precision with which a spin constraint might be inferred ultimately depends on our ability to calibrate or independently constrain the field properties with observations.

\subsection{Image-integrated \texorpdfstring{$\angle\beta_2$}{arg(beta2)}}

Finally, we evaluate how the image-integrated polarization mode angle $\angle\beta_2$ as inferred by an EHT-like measurement changes as a function of spin. We compare between three different models: time-dependent MAD GRMHD simulations from \koral (gray), the force-free split-monopole BZ solution evaluated with its drift velocity (dashed red), and our inflow midplane solutions (blue). Figure~\ref{fig:spin_grmhd_inflow_comparison} summarizes this comparison. As in Figure~\ref{fig:spin_library_comparison}, the inflow calculations assume moderate magnetizations, $5 \leq \tilde{\phi} \leq 7$, and adopt $\Omega_{\rm Field}$ from Equation~\ref{eqn:omega_field_fit}. The three models exhibit the same monotonic trend with an increasingly radial polarization at higher values of spin. As expected, while GRMHD images exhibit a polarization structure that resembles the BZ monopole, finite plasma inertia shifts the value of $\angle\beta_2$ away from the force-free limit, especially at spin where the monopole geometry winds up rapidly. The inflow model reproduces this mixed behavior and yields image-integrated $\angle\beta_2$ values that quantitatively track the GRMHD distributions across spin. We fit the mean value of $\angle \beta_2$ from the inflow model over spin using the form introduced in Equation~37~of~\citet{chael_2025_radsurvey},
\begin{align}
\left[ \angle \beta_2 \right]_{\rm fit} = 2\ \arctan\left( \dfrac{-C_0}{\left| \Omega_H \right|}\right) + C_1,
\end{align}
and find $C_0 = 0.14$ and $C_1 = 4.82\ {\rm deg}$, with the largest deviations occurring for high spin.

The observed ranges of $\angle \beta_2$ in EHT measurements of M87$^*$ across different years are shown in Figure~\ref{fig:spin_grmhd_inflow_comparison} as horizontal magenta, orange, and purple bands, and the observed range of $\angle \beta_2$ for \sgra is shown as a horizontal orange band \citep{eht_m87_7,eht_sgra_7,eht_m87_20172021}. In both cases, measurements are subject to uncertainties about Faraday rotation and time averaging, although note that the \sgra measurement attempts to correct for Faraday rotation and can be treated as a kind of average over many dynamical times, due to the length of the observation \citep{wielgus_2022_sgrapolalma,wielgus_2024_internalfaraday,eht_sgra_7}. Taking all measurements at face value and assuming emission originates on field lines connected to the black hole, our moderate-magnetization models suggest that both M87$^*$ and \sgra have low to intermediate spins. This inference depends on the accuracy of our calibrations of $\tilde{\phi}$ and $\Omega_{\rm Field}$, however. For \sgra, the observational uncertainties are broader, which makes the constraint less conclusive: the GRMHD distribution is $< 2 \sigma$ from the observed range and indeed the only surviving models from the EHT analysis have $\bhspin = 0.9375$. The discrepancy between the multiple years of data for M87$^*$ could indicate the presence of a changing Faraday screen, and correcting for the Faraday rotation may shift the intrinsic $\angle \beta_2$ toward values more consistent with higher spins.\footnote{Astrophysical jet-power constraints infer that the M87$^*$ black hole has non-zero spin, although \citet{eht_m87_8} shows that intermediate spins of $\bhspin = 0.5$ are consistent with the constraint.} More generally, however, establishing a robust spin constraint requires addressing several systematic uncertainties, which we discuss in Section~\ref{sec:spin_constraints}.

Collectively, our findings establish the inflow model framework as both a reliable surrogate for full numerical simulations and a practical tool for probing departures from the standard simulation paradigm without incurring the cost of time-dependent GRMHD runs. More fundamentally, they show that $\angle \beta_2$ is a geometric tracer of spin anchored by the boundary condition imposed at the event horizon and subsequently reshaped by plasma dynamics in the near-horizon region. In this way, $\angle \beta_2$ provides a direct and measurable link between black hole spin, magnetospheric structure, and the polarized signatures accessible to observation.

\section{Discussion}
\label{sec:discussion}

Polarimetric images of black holes provide a direct probe of the electromagnetic structure near the event horizon, encoding information about magnetic field topology, fluid dynamics, black hole spin, and the flow of electromagnetic energy in the black hole magnetosphere.   To interpret these signatures in a physically transparent and computationally efficient way, we have introduced a semi-analytic inflow model for magnetized accretion. Our framework extends the split-monopole Blandford-Znajek solution by incorporating finite plasma inertia through a free magnetization parameter. It solves a reduced, stationary set of the GRMHD equations in the equatorial plane, with dynamics constrained by regularity at the fast magnetosonic point. The model is specified by four parameters: the magnetic flux $\tilde{\phi}$, the field line rotation rate $\Omega_{\rm Field}$, the location of the outer boundary $r_{\rm bdd}$, and the black hole spin.

We have applied the inflow model to investigate the physical origin of the spin dependence observed in polarimetric signatures of black hole accretion flows. In both GRMHD and the inflow model, higher black hole spin enhances magnetic winding and drives the field geometry toward a more azimuthal configuration, producing polarization patterns that are increasingly radially aligned. By tuning the magnetization and field-line rotation rate, we show that the inflow model quantitatively reproduces the spin-dependent behavior of the polarimetric angle $\angle \beta_2$ seen in GRMHD simulations, confirming that this angle serves as a direct proxy for magnetic winding and encapsulates the coupled influence of plasma dynamics and electromagnetic structure determined by the black hole spacetime.

The inflow model also clarifies why GRMHD simulations deviate from force-free expectations: finite plasma inertia alters the velocity and field structure in the accretion flow. By explicitly incorporating this effect, the inflow model provides a physically grounded method to interpolate between the idealized force-free limit and the fully turbulent GRMHD regime. By adjusting the magnetic flux and field line rotation rate, our model can self-consistently recover the structure of the electromagnetic field and trends in velocity as measured in simulations. In this way, it functions as a lightweight framework for exploring parameter space around GRMHD solutions and serves as a powerful interpretive tool for current and future polarimetric observations of black holes. In principle the model could be directly fit to data in the future to constrain both the plasma properties and black hole spin.

\subsection{Summary of results}

To assess how well the inflow model can emulate the results of time-dependent numerical simulations, we compared the inflow solutions to the emissivity-weighted structure of a GRMHD-simulated MAD accretion flow around an intermediate-spin black hole. We found that the plasma in the synchrotron-emitting region of the simulation was only moderately magnetized, and the field-line rotation rate was smaller than the force-free value, signifying a systematic departure from the high-magnetization regime. Despite these differences, however, we found that the inflow model can reproduce the qualitative radial structure of the fluid and field variables. The main discrepancy between the models was in the radial velocity, where the cold inflow model predicts systematically faster infall than the numerical simulation. This different is unsurprising, since the assumption that the inflowing matter is cold precludes the thermal pressure that provides support in GRMHD simulation. 

We further reported how boundary conditions shape the behavior of the inflow model. Increasing the outer boundary radius $r_{\rm bdd}$ or decreasing the field-line angular velocity $\Omega_{\rm Field}$ enhances both radial infall and magnetic winding, yielding larger $\left|u^r\right|$ and larger magnetic field pitch angles. Conversely, increasing $\Omega_{\rm Field}$ or imposing a Keplerian rotation profile at progressively smaller outer boundaries raises the average fluid angular velocity. These interdependencies highlight a fundamental tradeoff of the cold inflow framework, i.e., that adjusting boundary parameters to match one observable property (e.g., the magnetic field pitch angle) often introduces discrepancies in another (e.g., the velocity profile). Nevertheless, we argued that the inflow model offers a controlled and physically interpretable framework for isolating the role of plasma inertia in shaping magnetospheric structure. Indeed, despite its simplicity, the inflow model reproduces the essential magnetospheric structure of GRMHD simulations with striking fidelity. When the field-line rotation rate is matched to values inferred from the emissivity-weighted regions of simulations and the magnetic flux is tuned to moderate values ($\tilde{\phi} = 5-7$), we showed that the inflow model reproduces both the geometry of the magnetic field and velocity profiles of the numerical simulations. 

The inflow model also reproduces the radial dependence of $\angle \beta_2$ in simulated black hole images. We introduced a simple linear fit to the GRMHD-derived field-line rotation rate,  $\Omega_{\rm Field} \approx 0.05\,\bhspin + 0.015$, which we used to extend the model across a broader range of spins. We found that the model quantitatively predicts the spin dependence of the polarimetric image, with radial profiles of $\angle \beta_2$ converging to a spin-dependent asymptotic value near the inner shadow boundary as expected and falling into distinct radial bands at larger radii. The image-integrated values of $\angle \beta_2$ from the inflow model also show good agreement with image-averaged values from GRMHD simulations, so long as the field line rotation rate and inflow magnetization are chosen consistent with the GRMHD simulations.   This provides a physically grounded, quantitative understanding of how $\angle \beta_2$ varies with spin, along with tunable parameters to explore deviations from the standard numerical-GRMHD-simulation paradigm. Taken together, we argue that these results establish $\angle \beta_2$ as a promising tracer of black hole spin.  The dominant systematic uncertainty is in the field line rotation rate and magnetization of the inflow: if the latter deviate from GRMHD MAD values, the uncertainty in relating $\angle \beta_2$ and spin is correspondingly larger (Fig. \ref{fig:beta2_vs_radius}).

\subsection{Observational constraints on spin}
\label{sec:spin_constraints}

By directly comparing the observed values of $\angle \beta_2$ to both the inflow model and our GRMHD simulation library, we can infer tentative constraint on the spin of M87$^*$. EHT observations in 2017 reported $-163^\circ < \angle \beta_2 < -129^\circ$ \citep{eht_m87_7}, and in 2018 and 2021, the ranges were $-156^\circ < \angle \beta_2 < -99^\circ$ and $156^\circ < \angle \beta_2 < 168^\circ$, respectively. While data from 2017 and 2018 are broadly consistent, the 2021 data exhibits a marked shift by $\sim -60^\circ$, which infers a combination of source variability and the presence of a varying Faraday screen, both of which would have to be corrected to be comparable to our modeling. Nevertheless, taking the intervals at face value, our models favor a low-to-intermediate black hole spin.
We emphasize, however, that Faraday rotation and variability remain major uncertainties in this interpretation, and repeated observations across multiple frequencies and over multiple years will be essential in determining the robustness of the constraint.

It is also possible to infer a spin constraint for \sgra, however, this too is complicated by systematic uncertainties due to the measurement of potentially significant Faraday rotation along the line of sight. In contrast to M87$^*$, though, EHT observations of \sgra correspond to many dynamical times and may better represent the averaged state of the system. Assuming the presence of an external Faraday screen, derotating the observations of \sgra from 2017 infer the range $-168^\circ < \angle \beta_2 < -85^\circ$ \citep{eht_sgra_7,eht_sgra_8}. Although this wider interval is less constraining than in the case of M87$^*$, it remains broadly consistent with low- to intermediate-spin inflow models, provided that the external screen interpretation is correct.

These uncertainties highlight the importance of multi-frequency polarimetric observations. Higher-frequency data will suffer less from Faraday effects while probing regions closer to the event horizon, where spin signatures are most pronounced. In addition, spatially resolved rotation-measure maps, combined with dynamical modeling of the Faraday screen, can help disentangle propagation effects from the intrinsic geometric signatures considered here. Time-averaged images from data across multiple epochs will also be crucial, since short-term variability can obscure the underlying spin-dependent trends, especially in the case of M87$^*$. At the same time, refining the set of viable theoretical models, for example by focusing on the subset of GRMHD simulations consistent with observations (decreasing the scatter of the GRMHD in Figure~\ref{fig:spin_grmhd_inflow_comparison}), will sharpen theoretical priors and improve the precision of spin inference. We therefore expect that upcoming observations, together with improved theoretical constraints, will enhance our ability to isolate the intrinsic polarimetric structure and strengthen constraints on black hole spin.

\subsection{Limitations and outlook}

Although the inflow model reproduces many of the key trends observed in MAD GRMHD simulations, it is subject to several important limitations. By construction, the framework is restricted to stationary, axisymmetric, midplane solutions and therefore neglects turbulence, variability, and off-midplane emission, all of which may feature prominently in real accretion flows.
Similarly, the inflow framework does not capture the full complexity of the accretion environment, which may include jet-launching regions, large-scale outflows, and other nonaxisymmetric structures that can leave distinct imprints on the polarimetric signal. Alternative accretion geometries, such as tilted or retrograde disks or flows fed by stellar winds, may also have different magnetic field configurations. Magnetic field geometries or strengths that differ significantly from GRMHD MAD values would increase the $\angle \beta_2$ degeneracy between $\bhspin$ and $\tilde{\phi}$.
It will therefore be necessary to compare the model to a broader set of GRMHD simulations that span these diverse regimes to assess how reliably the model can be used to interpret polarimetric data.

The model further assumes negligible (or perfectly corrected) Faraday rotation, but both internal and external screens can significantly distort polarization signatures. Additional, unmodeled Faraday effects may arise from the intrinsic structure of the accretion flow or from changes in plasma composition: for instance, helium accretion alters the thermodynamics and depolarization properties, and pair-enriched plasmas can suppress or enhance Faraday rotation and modify the degree of circular polarization. It is worth recognizing that we have only considered model comparisons to MAD accretion flows; this regime is naturally aligned with the inflow framework, since large magnetic flux threads the horizon and strongly regulates the plasma dynamics in the MAD regime. By contrast, the model is less applicable to SANE flows, which feature weaker magnetic fields, higher plasma densities, and stronger turbulence. In such cases, Faraday rotation is enhanced and a significant fraction of the polarized emission may arise from regions decoupled from the horizon-threaded magnetosphere, which might lead to qualitatively different polarimetric signatures.

Several extensions of the inflow model remain open for future work. In this paper, we focused on prograde, cold, magnetically dominated flows, but incorporating internal energy (and thermal pressure support) would refine the quantitative connection between spin and polarization under more general plasma conditions. Another interesting model extension would treat retrograde accretion. While we have assumed that the angular momentum of the black hole and the surrounding flow are aligned, there is no compelling reason to exclude counterrotating or tilted configurations; indeed, retrograde accretion has been observed in stellar disks \citep{young_2020_counterrotation} and is not unexpected in galactic nuclei. Retrograde flows have already been extensively explored in the context of black hole imaging, with GRMHD and polarimetric studies documenting distinct signatures in both the emission morphology and the polarization structure \citep{palumbo_2020_beta2,eht_m87_7,eht_sgra_8,chael_2025_radsurvey}. Our inflow model does not naturally cover the misaligned case, however, since it sets the magnetic field line rotation rate to the disk angular velocity (disallowing retrograde configurations) and assumes an equatorial flow (disallowing tilted configurations).

It would also be valuable to couple the inflow model to an outflow prescription in order to study the polarization signatures of the full accretion system. Recent work by \citet{gelles_2025_spinjetpol} has demonstrated that the polarimetric structure of jet emission can be used to locate the light cylinder, which in turn provides a constraint on black hole spin. By integrating an outflow component into our model, it would be possible to obtain a more complete description of how the near-horizon dynamics connects to energy extraction and imprints on the global accretion structure.   

Future VLBI campaigns with improved resolution, dynamic range, and multifrequency coverage will be essential for testing whether polarized emission originates on magnetic field lines that thread the event horizon or are instead anchored farther out in the accretion flow. Distinguishing between these scenarios will reveal whether jets are powered primarily by black hole spin or disk rotation, while clarifying how polarimetric structures encode the properties of the central black hole. Resolving these questions will tie polarization signatures directly to the physics of relativistic jet launching and provide the strongest test yet of spin as the engine of black hole energy extraction on galactic scales.

\begin{acknowledgements}
The authors thank Boris Georgiev, Zack Gelles, Nick Kokron, and Angelo Ricarte for helpful comments. G.N.W.\,was supported by the Taplin Fellowship and the Princeton Gravity Initiative. A.C.\,was supported by the Princeton Gravity Initiative. A.L.\,was supported in part by NSF grant 2307888. E.Q.\,was supported in part by a Simons Investigator award from the Simons Foundation.  This work benefited from EQ and AL's stay at the Aspen Center for Physics, which is supported by National Science Foundation grant PHY-2210452.
\end{acknowledgements}

\appendix

\section{Blandford-Znajek Monopole Detail}
\label{app:bzmodel}

We here review the force-free split monopole solution of \citet{blandford_1977_bz} in detail. We began with the perturbative expansion of the solution to subleading order in spin as presented in \citet{armas_2020_bzexpansion}. In terms of the three functions $\psi(r, \theta)$, $\Omega(\psi)$, and $I(\psi)$, the leading terms in the expansion are 
\begin{align}
    \psi =&\, \psi_0 \, (1-\cos{\theta})+\bhspin^2\,\psi_0\,\hat{R}\left(r\right)\,\sin^2\theta\,\cos\theta, \\
    \Omega =&\, \dfrac{\bhspin}{8} + \bhspin^3 \, \omega_2(\theta), \\
    I =&\, -2\pi\psi_0 \, \left[ \left(\dfrac{\bhspin}{8}\sin^2\theta\right) \right. + \nonumber \\
    &\, \left. \bhspin^3\, \sin^2 \theta \left( \omega_2(\theta) + \dfrac{1}{4} \hat{R}\left(r\right)\cos^2\theta\right) \right],
\end{align}
with $\hat{R}(x)$ and $\omega_2(\theta)$ given in Eqs.~(4.49) and (4.71) as
\begin{align}
    \hat{R}(x) =& \frac{1}{72x}\bigg[24+11x+36x^2-36x^3 + \nonumber \\
    & \left(6x + 18x^3 -36x^3\right)\log\left(\dfrac{x}{2}\right) + \nonumber \\
    & \left(27x^3 - 18x^4\right)\log\left(\dfrac{x}{2}\right)\log\left(\dfrac{x-2}{x}\right) + \nonumber \\
    & 9x^3 - 3 + 2 x \, \mathrm{Li}_2 \dfrac{2}{x}\bigg] \\
    \omega_2(\theta) =& \dfrac{1}{32}-\dfrac{4\hat{R}(2)-1}{64}\sin^2\theta, \\
    \hat{R}(2) =& \dfrac{6\pi^2-49}{72} = 0.14191 \ldots
\end{align}

Unfortunately, the perturbative expansion only satisfies the Znajek condition order by order in spin, and solving for the full solution is intractable. We obtain our approximate solution by matching boundary conditions for the field at the horizon and at infinity and selecting an ansatz for the behavior at intermediate radii. On the horizon, the Znajek condition enforces regularity. Applied to Equation~\ref{eqn:ansatz}, the Znajek condition is
\begin{align}
\label{eqn:znajek}
I(\psi) = 2\pi(\Omega - \Omega_H) \frac{(r_+^2 + \bhspin^2)\sin\theta}{r_+^2 + \bhspin^2\cos^2\theta} \, \partial_\theta \psi,
\end{align}
where $\Omega_H = \bhspin / (2 M r_+)$ is the angular velocity of the black hole horizon. In the large-$r$ limit, the electromagnetic field reduces to the well-known flat-space force-free monopole to third order in spin. To leading order in perturbation theory, the magnetic flux then takes the form
\begin{align}
\psi(r,\theta) = 1 - \cos\theta,
\end{align}
and the current and rotation rate satisfy
\begin{align}
I(\psi) = -2\pi \, \psi \, \Omega(\psi) \, (2 - \psi), \quad \Omega(\psi) = \dfrac{\Omega_H}{2}.
\end{align}
Combining these expressions yields
\begin{align}
\label{eqn:ff_infty}
I(\psi) = - 2\pi \, \Omega(\psi) \sin\theta \, \partial_\theta \psi.
\end{align}

To interpolate between the two limits, we first define the ratio
\begin{align}
X(r,\theta) \equiv \dfrac{I}{2\pi(\Omega - \Omega_H) \, \partial_\theta \psi}.
\label{eqn:app:Xdefn}
\end{align}
In terms of $X$, then Znajek condition implies that
\begin{align}
X(r_+,\theta) = \frac{(r_+^2 + \bhspin^2)\sin\theta}{r_+^2 + \bhspin^2\cos^2\theta},
\end{align}
while the asymptotic behavior in Equation~\ref{eqn:ff_infty} implies that
\begin{align}
X(r \to \infty, \theta) = \sin\theta.
\end{align}
Any ansatz for $X(r,\theta)$ should reproduce both of these limits and provide a smooth interpolation across intermediate radii. We choose to take 
\begin{align}
X_{\rm approx}(r,\theta) &= \frac{\sqrt{\Pi(r,\theta)} \, \sin\theta}{\Sigma(r,\theta)}.
\end{align}

\begin{figure*}[htp!]
\centering
\includegraphics[width=\linewidth]{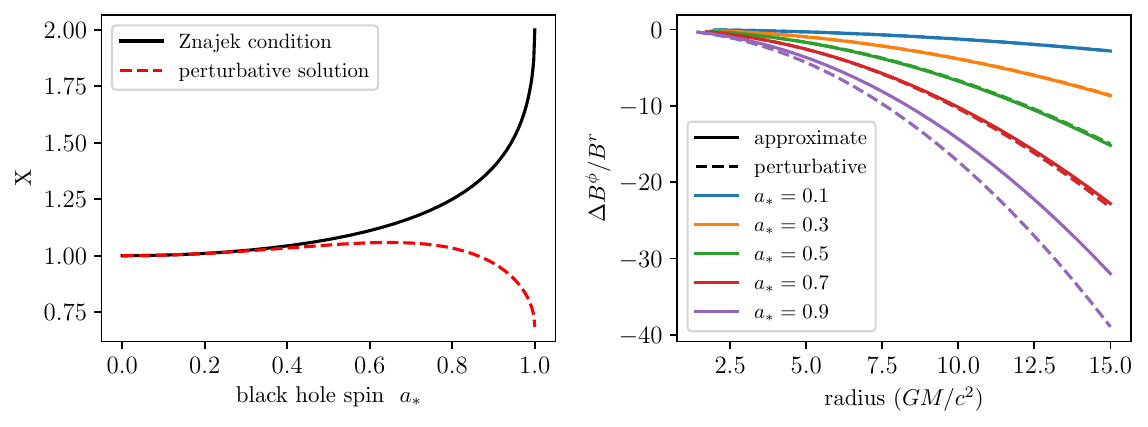}
\caption{
Comparison between approximation and perturbative Blandford-Znajek split monopole solution. The left panel shows how well the perturbative solution agrees with the Znajek condition in the midplane at the event horizon, expressed in terms of the ratio $X$ defined in Equation~\ref{eqn:app:Xdefn} (or equivalently Equation~\ref{eqn:X1}). The right panel shows the magnetic field winding ratio $\Delta B^\phi/B^r$ as a function of radius in the midplane. The ratio $X$ for the perturbative solution begins to disagree noticeably with the Znajek condition at intermediate values of spin.
}
\label{fig:app_comparison}
\end{figure*}

Figure~\ref{fig:app_comparison} shows a comparison between the approximate solution and the monopole solution in terms of both (left) the value of $X$ in the midplane at the event horizon and (right) the computed magnetic winding ratio $\Delta B^\phi / B^r$ as a function of radius in the midplane. The construction of the approximate solution guarantees that $X$ agrees with the Znajek condition at the horizon, whereas the perturbative solution may exhibit significant variations from this value. Despite disagreement in $X$ beginning at intermediate values of $\bhspin$ (and not just at the event horizon), the field winding ratio typically agrees quite well between the perturbative and approximate solutions at small radius and up to large spin.

\section{Inflow Model Detail}
\label{app:inflow_model}

Our inflow model yields a cold, equatorial, stationary, and axisymmetric ideal MHD solution for a magnetized plasma accreting onto a Kerr black hole. The evolution is confined to the equatorial plane and governed by a subset of the full GRMHD system. 

\subsection{Governing equations and constraints}

The model has six dynamical variables
\begin{align}
\rho,\quad u^t,\quad u^r,\quad u^\phi,\quad F_{t\theta},\quad F_{\theta\phi},
\end{align}
which are constrained by six conserved or algebraically related quantities
\begin{align}
\mathcal{F}_M &= 2\pi r^2 \rho u^r, \\
\mathcal{F}_E &= -2\pi r^2 {T^r}_t, \\
\mathcal{F}_L &= 2\pi r^2 {T^r}_\phi, \\
\Phi_B &= \pi F_{\theta\phi},
\end{align}
representing mass flux, energy flux, angular momentum flux, and magnetic flux, respectively, along with a fixed field line rotation rate $\Omega_F,$ and the condition that $-u_\mu u^\mu = 1.$

Our restriction to the cold plasma limit sets $u = P = 0$, so that the stress-energy tensor becomes
\begin{align}
T^{\mu\nu} = \rho u^\mu u^\nu + 
F^{\mu\alpha} {F^\nu}_\alpha - \frac{1}{4} g^{\mu\nu} F^{\alpha\beta} F_{\alpha\beta}.
\end{align}

Our model lives in the equatorial plane $\theta = \pi/2$, and we assume that the poloidal motion and magnetic field are purely radial, i.e., $u^\theta = B_\theta = 0$. This implies that $F_{r\phi} = 0$. Under the ideal MHD condition,
\begin{align}
u^\mu F_{\mu\nu} = 0,
\end{align}
combined with the antisymmetry of $F_{\mu\nu}$, we find that
\begin{align}
u^t F_{tr} + u^\phi F_{\phi r} &= 0 \quad \Rightarrow \quad F_{tr} = 0, \\
u^t F_{t\phi} + u^r F_{r\phi} &= 0 \quad \Rightarrow \quad F_{t\phi} = 0, 
\end{align}
so that the only nonzero components of $F_{\mu\nu}$ are $F_{t\theta}, F_{r\theta}, F_{\theta\phi}$. The relationship between the remaining nonzero components can be determined by the ideal MHD constraint $u^\mu F_{\mu\nu} = 0$ so that
\begin{align}
F_{r\theta} = -\frac{F_{t\theta} u^t - F_{\theta\phi} u^\phi}{u^r}.
\end{align}

Assuming time stationarity ($\partial_t \to 0$) and axisymmetry ($\partial_\phi \to 0$), the additional requirement of electromagnetic degeneracy ($\hodgestar F_{\mu\nu} F^{\mu\nu} = 0$) allows us to relate $F_{t\theta}$ and $F_{\theta\phi}$ via the law of isorotation. Recall that the Faraday tensor $F_{\mu\nu} = \partial_\mu A_\nu - \partial_\nu A_\mu$, so that degeneracy implies that the gradients of $A_t$ and $A_\phi$ are aligned, 
\begin{align}
(\partial_\theta A_\phi) \, ( \partial_r A_t ) - (\partial_\theta A_t ) \, (\partial_r A_\phi) = 0.
\end{align}
This ensures that there exists some scalar function $\Omega_F(r,\theta)$ such that
\begin{align}
\dfrac{\partial_r A_t}{\partial_r A_\phi} = \dfrac{\partial_\theta A_t} {\partial_\theta A_\phi} = -\Omega_F.
\end{align}

This function $\Omega_F$ defines the angular velocity of field lines, and the law of isorotation then follows:
\begin{align}
F_{t\theta} &= \partial_t A_\theta - \partial_\theta A_t \\
&= -\partial_\theta A_t \\
&= \Omega_F \, \partial_\theta A_\phi \\
&= \Omega_F F_{\theta\phi}. \label{eqn:app:isorotation}
\end{align}

At the radius $r_{\rm bdd}$ where $u^r = 0$, conservation of energy and angular momentum enable computation of a direct relationship between the energy flux $\mathcal{F}_E$, angular momentum flux $\mathcal{F}_L$, and the mass flux $\mathcal{F}_M$. Multiplying the angular momentum flux by $\Omega = u^\phi / u^t$, we find
\begin{align}
\Omega \, \mathcal{F}_L &= 2\pi r^2 \left( \rho u^r u_\phi \, \Omega + F^{r\theta} F_{\phi\theta} \, \Omega \right) \\
&= 2\pi r^2 \left( \rho u^r u\phi \Omega - F^{r\theta} F_{t\theta} \right),
\end{align}
where we have used the law of isorotation and the fact that at $r_{\rm bdd}$, $u^r \to 0$ and the fluid moves purely azimuthally so that flux freezing implies $\Omega_F = \Omega = u^\phi / u^t$, i.e., the angular velocity of the field lines is equal to the fluid angular velocity.

By including a contribution from mass flux
\begin{align}
(-u_t - u_\phi \Omega)\, \mathcal{F}_M = 2\pi r^2 \rho u^r (-u_t - u_\phi \Omega)
\end{align}
and substituting in for $\mathcal{F}_L$, we find that
\begingroup
\small
\begin{align}
\mathcal{F}_E &= \Omega \mathcal{F}_L + (-u_t - u_\phi \Omega) \mathcal{F}_M \\
&= 2\pi r^2 \left( \rho u^r u_\phi \Omega - F^{r\theta} F_{t\theta} - \rho u^r \left(u_t + u_\phi \Omega\right) \right) \\
&= 2\pi r^2 \left( -\rho u^r u_t -F^{r\theta} F_{t\theta} \right) \\
&= \mathcal{F}_E.
\end{align}
\endgroup
This identity conveniently relates all conserved fluxes at the outer boundary and allows $\mathcal{F}_E$ to be computed algebraically once $\mathcal{F}_L$ is known.

\subsection{Input parameters and closure}

The inflow solution is fully determined once a minimal set of physical parameters are specified. We choose the following:
\begin{itemize}[leftmargin=*]
    \item The mass flux $\mathcal{F}_M$, which we normalize to $-1$,
    \item The location of the outer boundary, which is subject to the condition $u^r(r_{\rm bdd}) = 0$,
    \item The field line angular velocity $\Omega_F$, specified through the fluid velocity $\Omega = u^\phi/u^t$ at $r_{\rm bdd}$, and
    \item The magnetic flux in the equator, encoded by $F_{\theta\phi}$ or equivalently $\tilde{\phi} = F_{\theta\phi}\sqrt{\pi}/2$. \\
\end{itemize}

Together with $u_\mu u^\mu = -1$, these constraints fix all but one degree of freedom. The remaining parameter is determined by requiring regularity of the solution at the fast magnetosonic point $r = r_{\rm fast}$, where the magnetosonic wave speed matches the flow velocity and beyond which information cannot propagate upstream.

\begin{figure*}[htp!]
\centering
\includegraphics[width=\linewidth]{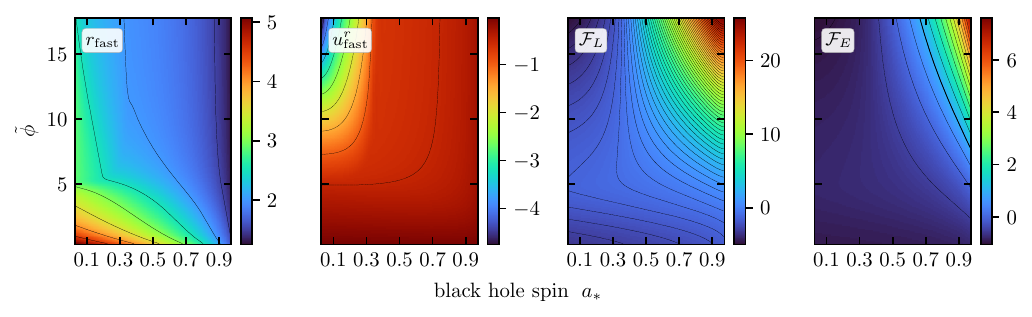}
\caption{
Eigenvalues of inflow model solutions as a function of black hole spin and magnetization parameter $\tilde{\phi}$. All solutions have their outer radial boundary at the ISCO and use the Keplerian rotation rate for $\Omega_{\rm Field} = \Omega_{\rm Fluid} = u^\phi/u^t$ at that point. Panels show the fast point radius $r_{\rm fast}$, radial velocity at the fast point $u^r_{\rm fast}$, angular momentum flux $\mathcal{F}_L$, and energy flux $\mathcal{F}_E$. Contours are spaced by $0.5$ in each panel. In the $\mathcal{F}_E$ panel, the thick black contour denotes $\mathcal{F}_E = 1$, above which the efficiency exceeds $100\%$ and thus energy is extracted from the black hole.
}
\label{fig:app_seeds}
\end{figure*}

At the critical point, the differential equation governing the evolution of the velocity become singular unless the numerator of the expression also vanishes, and so for the solution to be physically admissible and smoothly transonic, the system must satisfy a regularity condition. Following \citet{gammie_1999_inflow}, we solve for regularity by satisfying a saddle point constraint on the energy flux $\mathcal{F}_E$, treated as a function of radius $r$, radial velocity $u^r$, and the unknown flux $\mathcal{F}_L$. In particular, we require that
\begin{align}
\label{eqn:fastpoint_conditions}
\partial_r \mathcal{F}_E(r, u^r, \mathcal{F}_L) &= 0, \\
\partial_{u^r} \mathcal{F}_E(r, u^r, \mathcal{F}_L) &= 0, \\
\mathcal{F}_E(r, u^r, \mathcal{F}_L) &= \mathcal{F}_E(r_{\rm bdd}, 0, \mathcal{F}_L),
\end{align}
where the final condition ensures energy conservation across radii by equating the value of $\mathcal{F}_E$ at the fast point to its value at the outer boundary $r_{\rm bdd}$.

\subsection{Numerical details}

We solve the root-finding problem numerically by evaluating $\mathcal{F}_E$ and its derivatives at each iteration, treating the other parameters as fixed. We use a globally convergent Newton-Raphson method with line search for the minimization problem. In certain regions of parameter space, the solution becomes difficult to track due to discontinuities in $\mathcal{F}_E$ arising from root-branch transitions.

Once the fast point is located, the full radial structure of the inflow is recovered by explicitly solving for $u^r$ at each radius. The other components of $u^\mu$ are reconstructed using the input $\Omega_F$ and algebraic relations derived from the ideal MHD condition. The full stress-energy tensor $T^{\mu\nu}$ and electromagnetic tensor $F^{\mu\nu}$ are then computed using standard expressions for cold, ideal GRMHD.

To explore inflow solutions across black hole spin and magnetic flux, we use a continuation method starting from known saddle point solutions and using previous solutions as initial guesses. Representative seed points for $( \bhspin, \tilde\phi ) \to ( r_{\rm fast}, u^r_{\rm fast}, \mathcal{F}_L )$ include
\begin{itemize}
\item $(0.5, 4) \quad \to \quad \left( 2.45, -0.399, -1.065 \right)$,
\item $(0.5, 15) \quad \to \quad \left( 1.899, -0.608, 2.949 \right)$,
\end{itemize}
where we take the radial boundary to be at the ISCO and set $\Omega$ to the Keplerian value at that point.
Figure~\ref{fig:app_seeds} shows some of the accessible domain in $(\bhspin, \tilde\phi)$ space generated using this stepping procedure.

\bibliography{main}

\end{document}